\documentclass[journal]{IEEEtran}

\usepackage{xcolor,soul,framed} 

\colorlet{shadecolor}{yellow}
\usepackage[pdftex]{graphicx}
\graphicspath{{../pdf/}{../jpeg/}}
\DeclareGraphicsExtensions{.pdf,.jpeg,.png}

\usepackage[cmex10]{amsmath}
\usepackage{array}
\usepackage{mdwmath}
\usepackage{mdwtab}
\usepackage{eqparbox}
\usepackage{url}

\usepackage[compress]{cite}
\usepackage{graphics}
\usepackage{subfig}
\usepackage{booktabs}
\usepackage[justification=centering]{caption}
\usepackage{newtxtext,newtxmath}

\begin{document}

\title{Local Geometric Distortions Resilient Watermarking Scheme Based on Symmetry}

\author{Zehua~Ma, Weiming~Zhang, Han~Fang, Xiaoyi~Dong, Linfeng~Geng, and Nenghai~Yu

\thanks{All the authors are with CAS Key Laboratory of Electromagnetic Space Information, University of Science and Technology of China, Hefei, 230026, China. (e-mail: mzh045@mail.ustc.edu.cn, zhangwm@ustc.edu.cn, ynh@ustc.edu.cn) Corresponding author: Weiming Zhang.}
\thanks{This work was supported in part by the Natural Science Foundation of China under Grant U1636201 and 61572452, and by Anhui Initiative in Quantum Information Technologies under Grant AHY150400.
}
}  

\maketitle


\begin{abstract}
As an efficient watermark attack method, geometric distortions destroy the synchronization between watermark encoder and decoder. And the local geometric distortion is a famous challenge in the watermark field. Although a lot of geometric distortions resilient watermarking schemes have been proposed, few of them perform well against local geometric distortion like random bending attack (RBA). To address this problem, this paper proposes a novel watermark synchronization process and the corresponding watermarking scheme. In our scheme, the watermark bits are represented by random patterns. The message is encoded to get a watermark unit, and the watermark unit is flipped to generate a symmetrical watermark. Then the symmetrical watermark is embedded into the spatial domain of the host image in an additive way. In watermark extraction, we first get the theoretically mean-square error minimized estimation of the watermark. Then the auto-convolution function is applied to this estimation to detect the symmetry and get a watermark units map. According to this map, the watermark can be accurately synchronized, and then the extraction can be done. Experimental results demonstrate the excellent robustness of the proposed watermarking scheme to local geometric distortions, global geometric distortions, common image processing operations, and some kinds of combined attacks.
\end{abstract}


\IEEEpeerreviewmaketitle


\section{Introduction}\label{introduction}

\IEEEPARstart{D}{igital} watermark is widely used for ownership protection, authentication, annotation, etc. \cite{history1,history2}. An efficient watermarking scheme should be robust to a variety of distortions. Besides the robustness towards common image processing, the robustness towards geometric distortions is equally significant. Common image processing, for example, filtering or compression, weakens the watermark by changing the value of pixels. However, geometric distortions destroy the synchronization between watermark encoder and decoder. Under such distortions, the decoder can no longer decode the watermark without resynchronization even through the watermark still exists.

The geometric distortions can be divided into two categories. One is global geometric distortion, such as rotation, scaling, translation, and cropping. The other is local geometric distortion, like random bending attacks (RBA) of Stirmark \cite{stirmark1,stirmark2}. The local geometric distortions are more complicated and harder to recover. And it has been an open problem for years that designing a watermarking scheme performing well under local geometric distortions. Besides, many cross-media watermarking schemes have been proposed recently, for example, print-scanning watermarking schemes \cite{ULPM,guo2007paired}, print-camera watermarking schemes \cite{fang2019camera,chen2016picode,chen2017ra}, and screen-camera watermarking schemes \cite{RU,Fanghan,lee2010digital,unseencode}. Considering that the cross-media process will introduce unnoticeable local geometric distortions \cite{stirmark1}, a watermark synchronization process resilient to local geometric distortions may improve the performance of such schemes.

A lot of geometric distortions resilient watermarking schemes have been proposed. And they can be divided into the following five categories. The first category is the watermarking scheme using image normalization to resist geometric transforms \cite{norm1,norm2,norm3,norm4,norm5}. By normalizing the image according to a set of predefined moment criteria, watermarked image achieves invariance under affine transformation. But image normalization based methods have weak performance under cropping distortions. Because cropping changes the moments calculated from the whole image. Meanwhile, most image normalization based methods have only one-bit capacity and high computational cost.

The second category is based on template \cite{template0,template1,template2}. A template is embedded in the Fourier domain. Before watermark extraction, the affine transformation can be estimated by comparing the detected template with the original one. Then the inverse affine transformation is implemented and the watermark can be extracted with a traditional method. However, the performance of template based watermark highly depends on the detection accuracy of the template. Besides, such watermark is vulnerable to template removal attack \cite{templateattack}.

The third category is the watermarking scheme embedding the watermark in the geometric invariant domain \cite{inv1,inv2,inv3,inv4,ULPM}. A rotation, scaling and translation (RST) invariant domain is obtained by applying the Fourier-Mellin transform (discrete Fourier transform and log-polar mapping). The RST operations in the spatial domain can be converted into parameter translation in the Fourier-Mellin domain. By embedding and extracting the watermark in the Fourier-Mellin domain, there is no need to estimate and invert the geometric distortions. And the translation can be easily recovered by using tracking sequences embedded along with the informative watermark. Kang \emph{et.al} \cite{ULPM} proposed a uniform log-polar mapping (ULPM) based watermarking scheme which eliminates the interpolation distortion and interference distortion introduced in the embedding process. Besides, Urvoy \emph{et.al} proposed a perceptual Discrete Fourier Transform (DFT) watermarking scheme \cite{urvoy2014perceptual}, whose embedding strategy can help these DFT-based watermarking schemes get perceptually optimal visibility.

\begin{figure}[t]
    \begin{center}
    \subfloat[The host image.]{
        {\centering\includegraphics[width=1.55in]{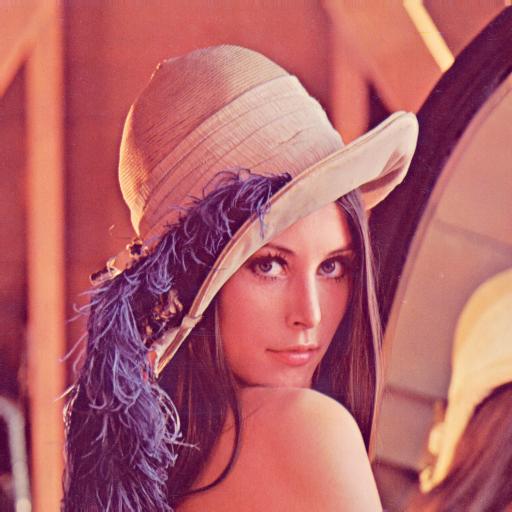}
        }
    }
    \subfloat[The image attacked by RBA.]{
        {\centering\includegraphics[width=1.55in]{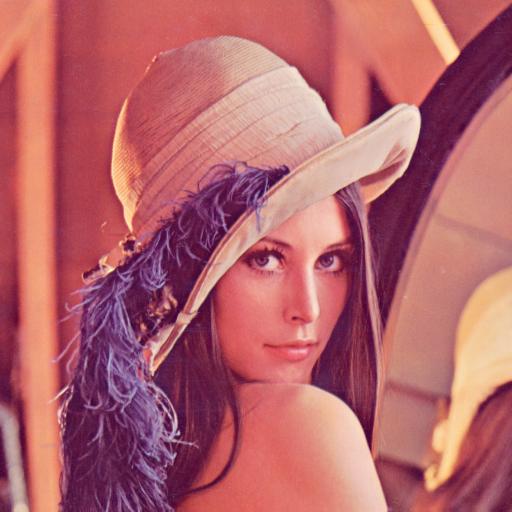}
        }
    }
    \\
    \subfloat[A regular grid.]{
        {\centering\includegraphics[width=1.55in]{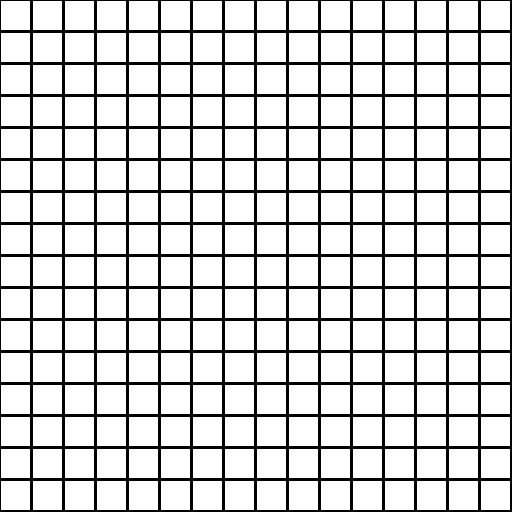}
        }
    }
    \subfloat[The grid attacked by the same RBA (red).]{
        {\centering\includegraphics[width=1.55in]{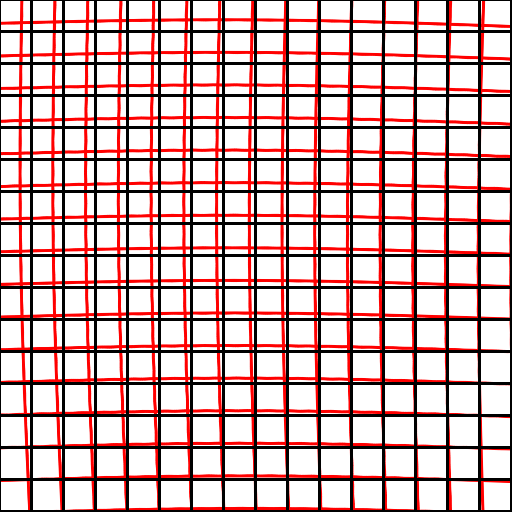}
        }
    }
    \end{center}
\caption{Two images attacked with the same RBA.}
\label{fig_RBA}
\end{figure}

However, local geometric distortion is still a challenge to the watermarking schemes belonging to the aforementioned three categories. Comparing Fig.~\ref{fig_RBA}(b) to Fig.~\ref{fig_RBA}(a), the distortion caused by RBA is almost unnoticeable. But the distortion is severe as shown in Fig.~\ref{fig_RBA}(d). Frequency values and moments which are calculated from the whole image will also change severely after RBA, which severely limits the performance of these watermarking schemes.

The feature based watermarking schemes belong to the fourth category \cite{feature1,feature2,feature3,feature4,feature5,feature6,LDFT}. By binding the watermark with the global or local geometrically invariant features, the watermarking scheme would obtain the corresponding robustness. Most of them are one-bit watermark scheme. In \cite{LDFT}, a multi-bit watermarking scheme was proposed based on local daisy feature transform (LDFT). It is resilient to many kinds of local geometric distortions. However, its robustness to common image processing operations and global geometric distortions is weak. A possible reason is that local geometrically invariant features are not robust enough to other distortions. Besides, comparing to other watermarking schemes, feature based methods usually have higher computational cost.

\captionsetup{font={footnotesize}}


The last category is the watermark scheme using periodic watermark itself as calibration signal \cite{acf1,acf2,acf3,acf4,Multibit}. A watermark unit is periodically tiled to generate a periodic watermark. In watermark extraction, the period can be detected by auto-correlation function (ACF) or magnitude spectrum (MS) and such period can act as a reference to recover the global geometric distortions. Considering that local geometric distortions can be regarded as a set of affine transformations in the local region, Voloshynovskiy \emph{et.al} \cite{Multibit} proposed using local ACF to detect and restore local geometric distortions. Moreover, as stated in the paper proposing template removal attack \cite{templateattack}, methods in this category are resilient to this attack.

In this paper, a watermark synchronization process based on symmetry is proposed, which is similar to the methods belonging to the last category mentioned above. However, we believe the difference between symmetry and periodicity makes symmetry have some advantages as the watermark calibration signal. And the corresponding watermarking scheme is proposed, which performs well under global geometric distortions, common image processing operations, and some combined attacks.

The rest of this paper is organized as follows. In Section~\ref{related_work}, a universal watermark synchronization process based on periodicity is illustrated first. Then the difference between it and the proposed synchronization process based on symmetry is discussed. In Section~\ref{embedding_process} and \ref{extraction_process}, we illustrate the watermark embedding process and the extraction process of the proposed watermarking scheme respectively. The experimental results are shown and discussed in Section~\ref{experiment}. Section~\ref{conclusion} draws the conclusion.

\section{Watermark Synchronization Process Analysis}\label{related_work}

\begin{figure*}[t]
    \begin{center}
    \subfloat[A example of watermark generating.]{
        {\centering\includegraphics[width=3.5in]{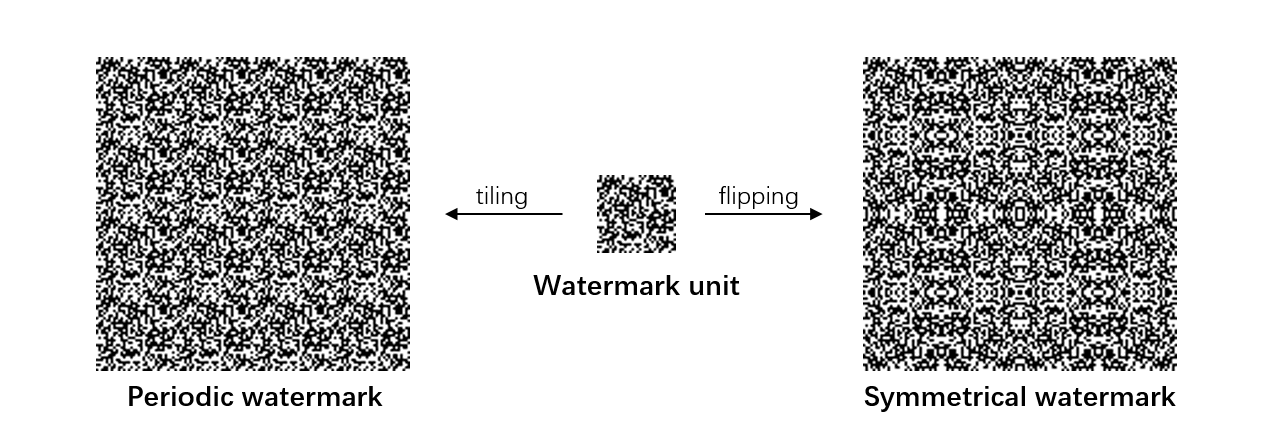}
        }
    }
    \subfloat[The corresponding state change of (a).]{
        {\centering\includegraphics[width=3.5in]{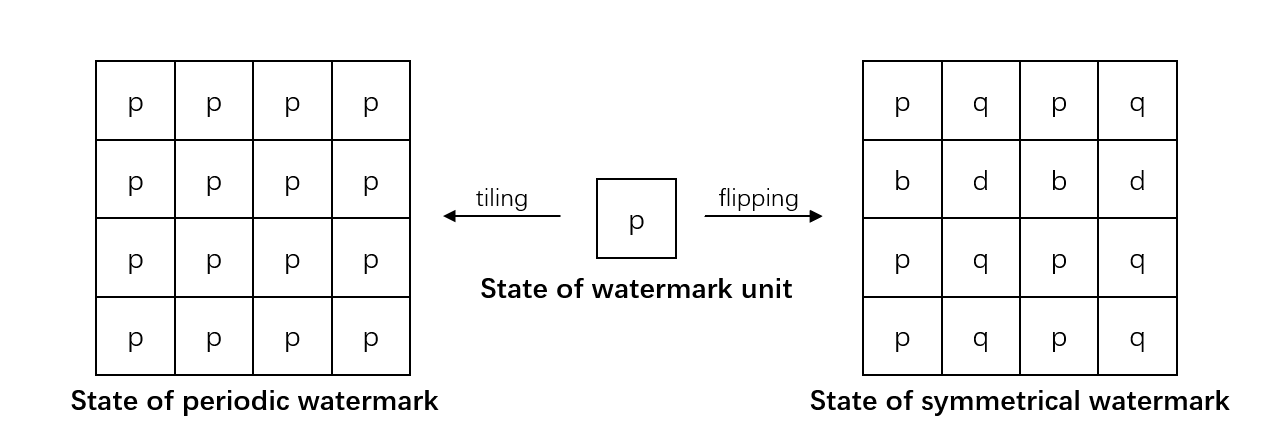}
        }
    }
    \end{center}
\caption{(a) is an example of generating a $4\times4$ periodic watermark by tiling a watermark unit and generating a $4\times4$ symmetrical watermark by flipping the same watermark unit. Using symbol `p' to represent the state of watermark unit, (b) is the corresponding state change of (a).}
\label{gen_sample}
\end{figure*}

\begin{figure*}[t]
  \begin{center}
  \includegraphics[width=0.9\textwidth]{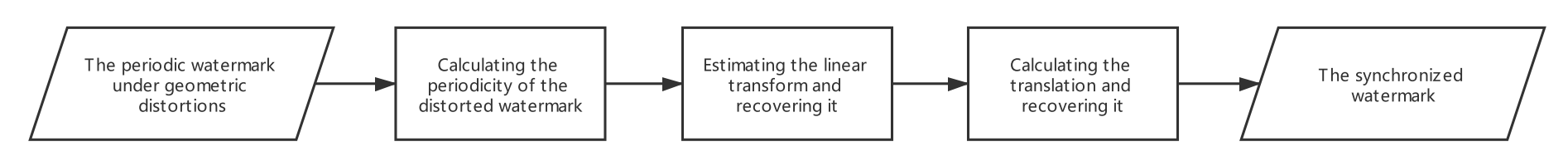}\\
  \caption{The flowchart of the universal watermark synchronization process based on periodicity.}\label{p_flowchart}
  \end{center}
\end{figure*}

\begin{figure*}[t]
  \begin{center}
  \includegraphics[width=0.9\textwidth]{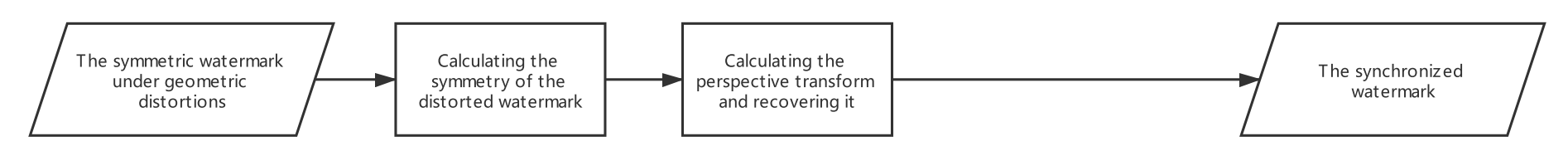}\\
  \caption{The flowchart of the proposed watermark synchronization process based on symmetry.}\label{s_flowchart}
  \end{center}
\end{figure*}


In this section, we will discuss and compare periodicity and symmetry, and their application on watermark synchronization to illustrate the advantages of symmetry as the watermark calibration signal.

\subsection{A Universal Periodic Watermark Synchronization Process}

The periodic watermark schemes, as mentioned in Section~\ref{introduction}, have been widely used. Among them, the scheme proposed by Voloshynovskiy \emph{et.al} \cite{Multibit} is one of the most representative schemes, as the unique periodic watermark scheme robust to local geometric distortions, or random bending attack. We first illustrate its watermark synchronization process to support the later discussion.

In \cite{Multibit}, the watermark unit, named macro-block, consisting of informative watermark and reference watermark. As shown in Fig.~\ref{gen_sample}(a), the watermark unit is tiled to generate a periodic watermark. In the case of rectangular macro-block, the watermark $\boldsymbol{W}$ has the following property:
\begin{equation}
    W(x,y)=W(x+m\ T_{len},y+n\ T_{wid})
\end{equation}
where $m$, $n$ are integers, and $T_{len}$, $T_{wid}$ are the size of the macro-block. The periodicity $\boldsymbol{P}$ of $\boldsymbol{W}$ is commonly defined as the auto-correlation of $\boldsymbol{W}$:
\begin{equation}\label{P_def}
    P(i,j)=\sum_{x} \sum_{y} W(x,y)W(x+i,y+j)
\end{equation}
in which $P(i,j)$ represents the correlation between $\boldsymbol{W}$ and the $\boldsymbol{W}$ translated by vector $\boldsymbol{(i,j)}$. $\boldsymbol{P}$ can also be computed by the frequency form of ACF:
\begin{equation}
    \boldsymbol{P}=IFFT[ FFT(\boldsymbol{W}) FFT(\boldsymbol{W})^*]
\end{equation}
where $FFT$ represents the fast Fourier transform, $IFFT$ is the corresponding inverse transform, and $*$ operator denotes complex conjugation. The periodicity will change along with the global geometric distortions.

The case of global geometric distortion is considered first. Most global geometric distortions can be uniquely described by an affine transformation, and the affine transformation can be represented by a linear component matrix $\boldsymbol{A}$, plus a translation component $\boldsymbol{v}$. Considering that the translation can be separately recovered, Voloshynovskiy \emph{et.al} \cite{Multibit} propose to use an approach based on penalized Maximum Likelihood (ML) estimation to estimate the linear component $\boldsymbol{A}$:
\begin{equation}\label{estimate_A}
\hat{\boldsymbol{A}}=\arg \min _{\boldsymbol{A} \in \Phi}\left\{\rho\left(\left(\begin{array}{l}
x^{\prime} \\
y^{\prime}
\end{array}\right)-\boldsymbol{A}\left(\begin{array}{l}
x \\
y
\end{array}\right)\right)+\mu \Omega(\boldsymbol{A})\right\}
\end{equation}
where $\hat{\boldsymbol{A}}$ is the optimal estimation within the set of possible solutions $\boldsymbol{A} \in \Phi$, $\rho$ denotes the cost function, and $(x,y)$, $(x^{\prime},y^{\prime})$ represent the Cartesian coordinates of ACF peaks before and after affine transformation. The last term $\mu \Omega(\boldsymbol{A})$ is a weighted prior knowledge, restricting the variations of parameters in $\boldsymbol{A}$. After recovering the linear transformation of the affine transformed watermark, the translation $\boldsymbol{v}$ can be easily recovered, for example based on an intercorrelation between the extracted watermark and the reference watermark mentioned above.

Considering that the local geometric distortions, like RBA and perspective transformation, can be approximated as affine transformations in the local region, and local geometric distortions can be approximated by a similar process. We can further summarize the watermark synchronization process proposed in \cite{Multibit} as the flowchart in Fig.~\ref{p_flowchart}.

It should be noted that even though flipping appeared in the scheme proposed by Voloshynovskiy \emph{et.al} \cite{Multibit} for some purposes, the synchronization process of the scheme is still based on periodicity and follows the above discussion.

\begin{figure*}[t]
    \begin{center}
    \subfloat[Original]{
        {\centering\includegraphics[width=.8in]{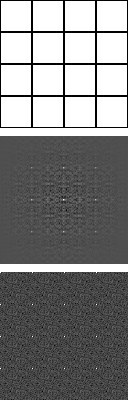}
        }
    }
    \subfloat[Rotation]{
        {\centering\includegraphics[width=.8in]{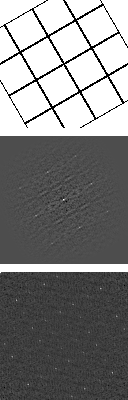}
        }
    }
    \subfloat[Scaling]{
        {\centering\includegraphics[width=.8in]{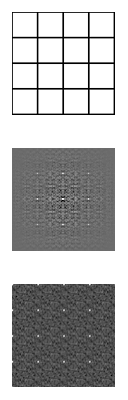}
        }
    }
    \subfloat[Translation]{
        {\centering\includegraphics[width=.8in]{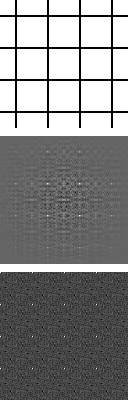}
        }
    }
    \subfloat[Cropping]{
        {\centering\includegraphics[width=.8in]{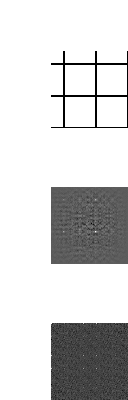}
        }
    }
    \subfloat[Affine transformation]{
        {\centering\includegraphics[width=.8in]{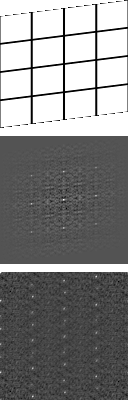}
        }
    }
    \subfloat[Perspective transformation]{
        {\centering\includegraphics[width=.8in]{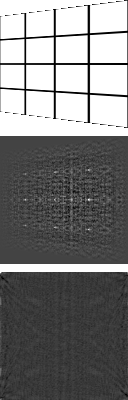}
        }
    }
    \end{center}
    \caption{\emph{Top row:} The grids under the same geometric distortions with watermarks, representing the states of the watermark units;\\ \emph{Middle row:}
The symmetry of the symmetrical watermark in Fig.~\ref{gen_sample}(a) under geometric distortions;\\ \emph{Bottom row:} The periodicity of the periodic watermark in Fig.~\ref{gen_sample}(a) under geometric distortions.}
    \label{samples_of_SP}
\end{figure*}

\subsection{The Proposed Symmetrical Watermark Synchronization Process}\label{illustration_of_SP}
However, we believe that the watermark synchronization process based on symmetry may have some advantages over the one based on periodicity. These advantages come from the difference between symmetry and periodicity and contribute to a simpler watermark synchronization process as shown in Fig.~\ref{s_flowchart}. Similarly, we define symmetrical watermark and its symmetry first. In our scheme, as shown in Fig.~\ref{gen_sample}(a), the symmetrical watermark $\boldsymbol{W}$ is generated by flipping the rectangular watermark unit, and it satisfies:
\begin{equation}
    W(x_s-x,y_s-y)=W(x_s+x,y_s+y)
\end{equation}
where $(x_s,y_s)$ are the coordinates of one symmetrical center of $\boldsymbol{W}$. In the symmetrical watermark $\boldsymbol{W}$, the symmetrical centers are the corners of the watermark unit. Similar to periodicity, the symmetry $\boldsymbol{S}$ of the watermark about a point $(i,j)$ can be defined by:
\begin{equation}\label{S1_def}
    S(i,j)=\sum_{x} \sum_{y} W(i-x,j-y)W(i+x,j+y).
\end{equation}
$S(i,j)$ is actually the correlation between $\boldsymbol{W}$ and the $\boldsymbol{W}$ flipped around point $(i,j)$.

We believe the most important difference between symmetry and periodicity is that there is a clearer mapping between symmetrical peaks and watermark position. Periodicity is a characteristic about translation, and symmetry is a characteristic about position. To be specific, in Eq.~(\ref{P_def}), the coordinates of $\boldsymbol{P}$ represent a translation or a vector, and the corresponding value in $\boldsymbol{P}$ denotes the possibility that the period of $\boldsymbol{W}$ is that vector. The coordinates of $\boldsymbol{P}$ and $\boldsymbol{W}$ have different meanings. But in Eq.~(\ref{S1_def}), the coordinates of $\boldsymbol{S}$ and $\boldsymbol{W}$ have the same meaning, indicating the position. Given a point $(i,j)$, the value of $S(i,j)$ represents the possibility that point $(i,j)$ is one of the symmetrical centers of $\boldsymbol{W}$. In the ideal case, the positions of symmetrical peaks in $\boldsymbol{S}$ are the positions of symmetrical centers in $\boldsymbol{W}$. 

Considering that symmetrical centers are corners of watermark units, knowing the position of symmetrical peaks is equivalent to knowing the position of watermark units and corresponding geometric distortions. Fig.~\ref{samples_of_SP} shows the periodicity of periodic watermark and the symmetry of symmetrical watermark under various geometric distortions, in which the watermarks in Fig.~\ref{gen_sample}(a) are used as samples and their periodicity and symmetry are calculated respectively by Eq.~(\ref{P_def}) and Eq.~(\ref{S1_def}). Observing these peaks in Fig.~\ref{samples_of_SP}, we find that symmetrical peaks show the position of the current watermark units and the experienced geometric distortions more clearly. Moreover, knowing the four corner points of a watermark unit, the position of the unit can be determined and most of its geometric distortions can be recovered, even including perspective transformation, just like Fig.~\ref{samples_of_SP}(g). So the watermark synchronization process based on symmetry shown in Fig.~\ref{s_flowchart} becomes simpler.  

As a summary of the above discussion, the advantages of the watermark synchronization process based on symmetry can be listed as follows:
\begin{enumerate}
    \item The process needs not the step of estimating translation. Comparing Fig.~\ref{samples_of_SP}(a) and Fig.~\ref{samples_of_SP}(d), symmetrical peaks show the position of watermark unit after translation while periodic peaks keep the same. Reducing the estimation steps contributes to a simpler synchronization process, which means less computational cost and higher accuracy. Meanwhile, the redundancy for estimating translation, like reference watermark, is no longer needed in the proposed watermark unit.
    \item The process performs better under local geometric distortions. In the proposed synchronization process, the geometric distortion on a watermark unit is regarded as a perspective transformation and will be recovered based on the detected four corners of the watermark unit. Meanwhile, as we have mentioned, the synchronization process based on periodicity uses affine transformations to approximate local geometric distortions, including perspective transformation and RBA. Comparing to affine transformation, we think that perspective transformation has more parameters and will fit better when approximating local geometric distortions.
\end{enumerate}

Besides, the proposed watermark synchronization process based on symmetry has the potential to become an improved version of the one based on periodicity. Generating the watermark by flipping rather than tiling and detecting the watermark by symmetrical peaks rather than periodic peaks, most periodic watermark schemes following the synchronization process in Fig.~\ref{p_flowchart} can be modified to the corresponding symmetrical watermark schemes. If the modification does not introduce new disadvantages, these modified watermark schemes will perform better.

\begin{figure*}[t]
  \begin{center}
  \includegraphics[width=0.9\textwidth]{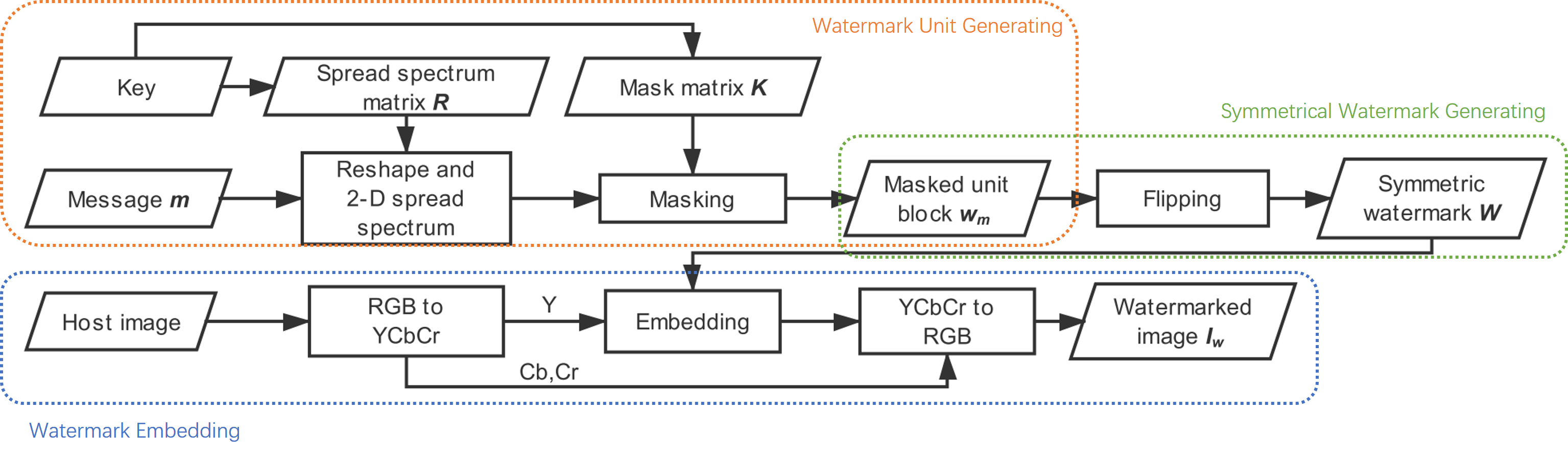}\\
  \caption{Watermark embedding process.}\label{embed}
  \end{center}
\end{figure*}

\begin{figure}
  \begin{center}
  \includegraphics[width=3.4in]{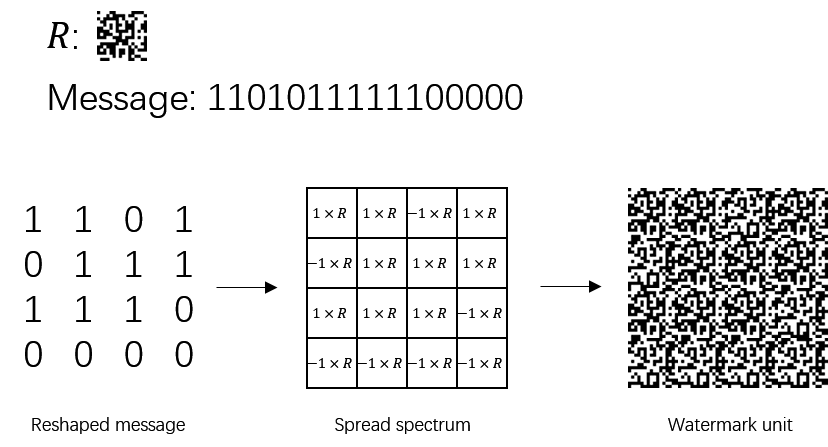}
  \caption{An example of using $\boldsymbol{R}$ to spread spectrum.}\label{2D_ss}
  \end{center}
\end{figure}

Actually, these disadvantages do exist and may limit the performance of synchronization process based on symmetry. These disadvantages can be listed as follows:
\begin{enumerate}
    \item The current process lacks a fast method to calculate symmetry. Calculating symmetry in space domain like Eq.~(\ref{S1_def}) is inefficient, especially when the host image is large. This problem also exists in the document watermark scheme proposed by Fang \emph{et.al} \cite{fang2019camera}.
    \item The flipping introduces additional states of the watermark unit, as shown in Fig.~\ref{gen_sample}(b), which need to be determined before decoding. A common solution is introducing predefined redundancy in the watermark as a reference. But such solutions will bring back the redundancy again, which is released from estimating translation.
\end{enumerate}

In Section~\ref{embedding_process} and \ref{extraction_process}, the proposed watermarking scheme based on symmetry is illustrated, which overcomes the disadvantages mentioned above. To be specific, we discover the relationship between symmetry and auto-convolution function, just like periodicity and auto-correlation function, and conclude a lower computational cost version of Eq.~(\ref{S1_def}). Meanwhile, we propose to use hypothesis testing to determine the state of the watermark rather than introducing redundancy. And the advantages of symmetry are retained in the proposed scheme to get better performance.

\section{Watermark Embedding}\label{embedding_process}

Fig.~\ref{embed} shows the framework of the proposed watermark embedding process. And it can be divided into three main modules: watermark unit generating, symmetrical watermark generating, and watermark embedding. The implementation details of these modules are illustrated as follows.

\subsection{Watermark Unit Generating}
By applying a key, we generate a 2-D bipolar random matrix $\boldsymbol{r}$ of size $L_{r}$, $\boldsymbol{r}_{i,j} \in$ \{-1,1\}, $i,j=1,2,...,L_{r}$. Then $\boldsymbol{r}$ is doubly upsampled to get $\boldsymbol{R}$. Comparing to $\boldsymbol{r}$, the components of $\boldsymbol{R}$ is mainly distributed in the middle frequency and more robust to common image processing operations. And the size of $\boldsymbol{R}$ is $L_{R}$, $L_{R}=2L_{r}$. Watermark sequence is first encoded using an error correction code (ECC) to obtain the message $\boldsymbol{m}$ of length $L_{m}$. Then we reshape $\boldsymbol{m}$ to a 2-D matrix of size $p \times q$, where $p \times q \geq L_{m}$. 
After that $\boldsymbol{m}$ is spread spectrum encoded using $\boldsymbol{R}$ to get a watermark unit $\boldsymbol{w}$. As Fig.~\ref{2D_ss} shows, in $\boldsymbol{w}$, bit `1' is presented as $+1 \times \boldsymbol{R}$ and bit `0' is presented as $-1 \times \boldsymbol{R}$. So the size of $\boldsymbol{w}$ is $L_{R}$ times the size of $\boldsymbol{m}$. Specifically, the length and width of $\boldsymbol{w}$ are defined as $m$ and $n$, where $m=p \times L_{R}$, $n=q \times L_{R}$. In the end, $\boldsymbol{w}$ is masked with a mask matrix $\boldsymbol{K}$, which is also generated from a key and doubly upsampled to the same size of $\boldsymbol{w}$. The masked watermark unit $\boldsymbol{w}_{m}$ is calculated by
\begin{equation}\label{mask_w}
    w_{m}(i,j)=w(i,j) K(i,j)
\end{equation}
where $i$, $j$ denote the index of the matrix. The masking operation is effective and necessary because it can
\begin{enumerate}
    \item Offer basic informative security. Without $\boldsymbol{K}$, even the watermarking scheme is a white box, the adversarial cannot accurately extract the message. 
    \item Eliminate the impact of weak messages. Similar to a weak key in cryptography, a weak message refers to a kind of message makes the watermark synchronization process behave in some undesirable ways. For example, a symmetrical 2-D message is a kind of weak message. Because the symmetry of the message itself and the symmetry generated by flipping will be both detected and confuse the watermark synchronization. To eliminate the impact of weak messages, watermark unit $\boldsymbol{w}$ is masked with a mask matrix $\boldsymbol{K}$. Because of spread-spectrum matrix, the information rate of $\boldsymbol{K}$ is $L_{R}^{2}$ times that of $\boldsymbol{w}$. The difference of information rate makes the properties of masked watermark unit $\boldsymbol{w}_{m}$ more depend on $\boldsymbol{R}$ rather than $\boldsymbol{w}$. So after masking, most properties of the message are eliminated, and $\boldsymbol{w}$ will be close to a random matrix. As a result, weak messages will no longer impact the synchronization process and the proposed watermarking scheme can reach its theoretical information rate.
    \item Help judge the state of the watermark unit. This part will be illustrated in Section~\ref{state_determined}.
\end{enumerate}


\subsection {Symmetrical Watermark Generating}\label{W_gen}
\begin{figure}
  \begin{center}
  \includegraphics[width=3.4in]{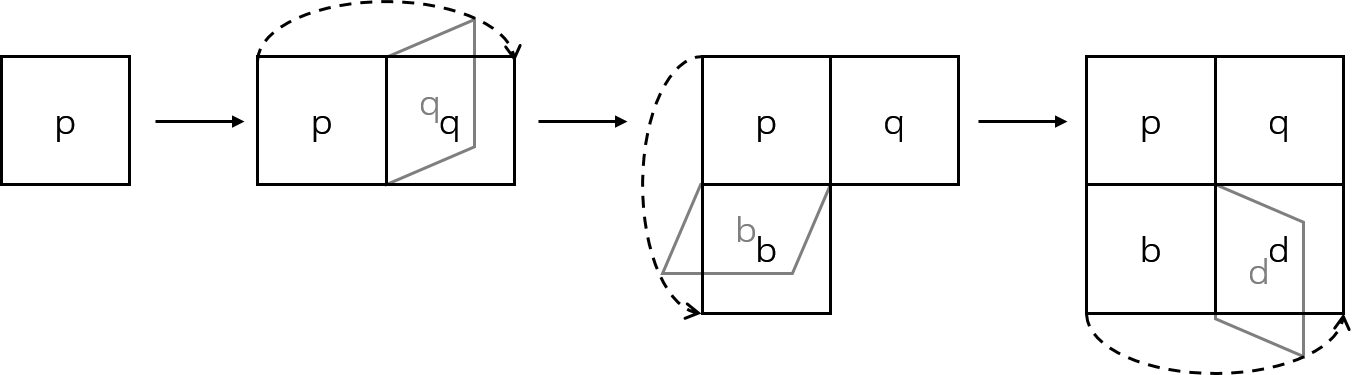}
  \caption{An example of generating symmetry by flipping watermark unit.}\label{flip}
  \end{center}
\end{figure}

We flip the masked watermark unit $\boldsymbol{w}_{m}$ to create the complete watermark $\boldsymbol{W}$. In this paper, flipping vertically is defined as flipping along the central horizontal axis of the block and flipping horizontally is defined as flipping along the central vertical axis of the block. Fig.~\ref{flip} shows an example of generating symmetry by flipping watermark unit in which we use symbol `p' to represent the watermark unit. The symmetrical watermark is generated by following a \emph{flipping rule}:
\begin{itemize}
    \item \emph{flipping rule}: The next horizontally adjacent watermark unit is generated by flipping the elder one horizontally, and the next vertically adjacent watermark unit is generated by flipping the elder one vertically.
\end{itemize}

To a specific image $\boldsymbol{I}$, $\boldsymbol{w}_m$ is flipped repeatedly until the size of $\boldsymbol{W}$ is larger than the size of $\boldsymbol{I}$. Then we crop $\boldsymbol{W}$ to the size of $\boldsymbol{I}$ to get the watermark for embedding.

Note that the flipping process is self-consistent. Different flipping order will generate the same symmetrical watermark $\boldsymbol{W}$.
To this $\boldsymbol{W}$, the adjacent line of two watermark units is the symmetrical axis of $\boldsymbol{W}$ and adjacent point of four watermark units is the symmetrical center of $\boldsymbol{W}$. 

\subsection {Watermark Embedding in Spatial Domain}
Watermark is embedded into the luminance of the host image with an additive way. In case that input is a color image, we will convert it to $YCbCr$ space and take component $Y$ as the luminance, called $\boldsymbol{I}$. To balance the robustness and imperceptibility, an adaptive watermark strength strategy is a common solution, for example, the adaptive embedding strategy proposed in \cite{huang2019enhancing}. In this paper, we use a simple strategy that embedding watermark with higher strength on the region with complex texture and with lower strength on the region with simple texture. The complexity of the image could be measured by the local variance of $\boldsymbol{I}$. So the adaptive watermark strength $\boldsymbol{s}$ is defined as follows:
\begin{equation}\label{adaptive strength}
    s(i,j)=\mathcal{F}(\sigma_{\boldsymbol{I}}^{2}(i,j))
\end{equation}
where $\boldsymbol{\sigma}_{\boldsymbol{I}}^{2}$ is the local variance of $\boldsymbol{I}$ and $\mathcal{F}$ is a nonlinear function. In the proposed scheme, $\mathcal{F}$ is defined as:
\begin{equation}
    \mathcal{F}(x)=\left\{\begin{array}{ll}{\alpha} & {, \quad if\quad log_2(x)<\alpha} \\ {log_2(x)} & {, \quad otherwise}\end{array}\right.
\end{equation}
where $\alpha$ is a global embedding strength and set to 2 in our scheme.
Then the watermarked image $\boldsymbol{I}_{w}$ is generated by
\begin{equation}
    I_{w}(i,j)=I(i,j)+s(i,j)W(i,j).
\end{equation}
The symmetrical watermark can be slightly pre-distorted to resist against spatial averaging and removal attack \cite{Multibit}. This predistortion will not significantly affect the symmetry of the watermark used for the recovering of geometric distortions.
In the end, a $YCbCr$ to $RGB$ transform will be applied if the host image is a color image.


\section{Watermark Extraction}\label{extraction_process}
\begin{figure}
  \begin{center}
  \includegraphics[width=3.2in]{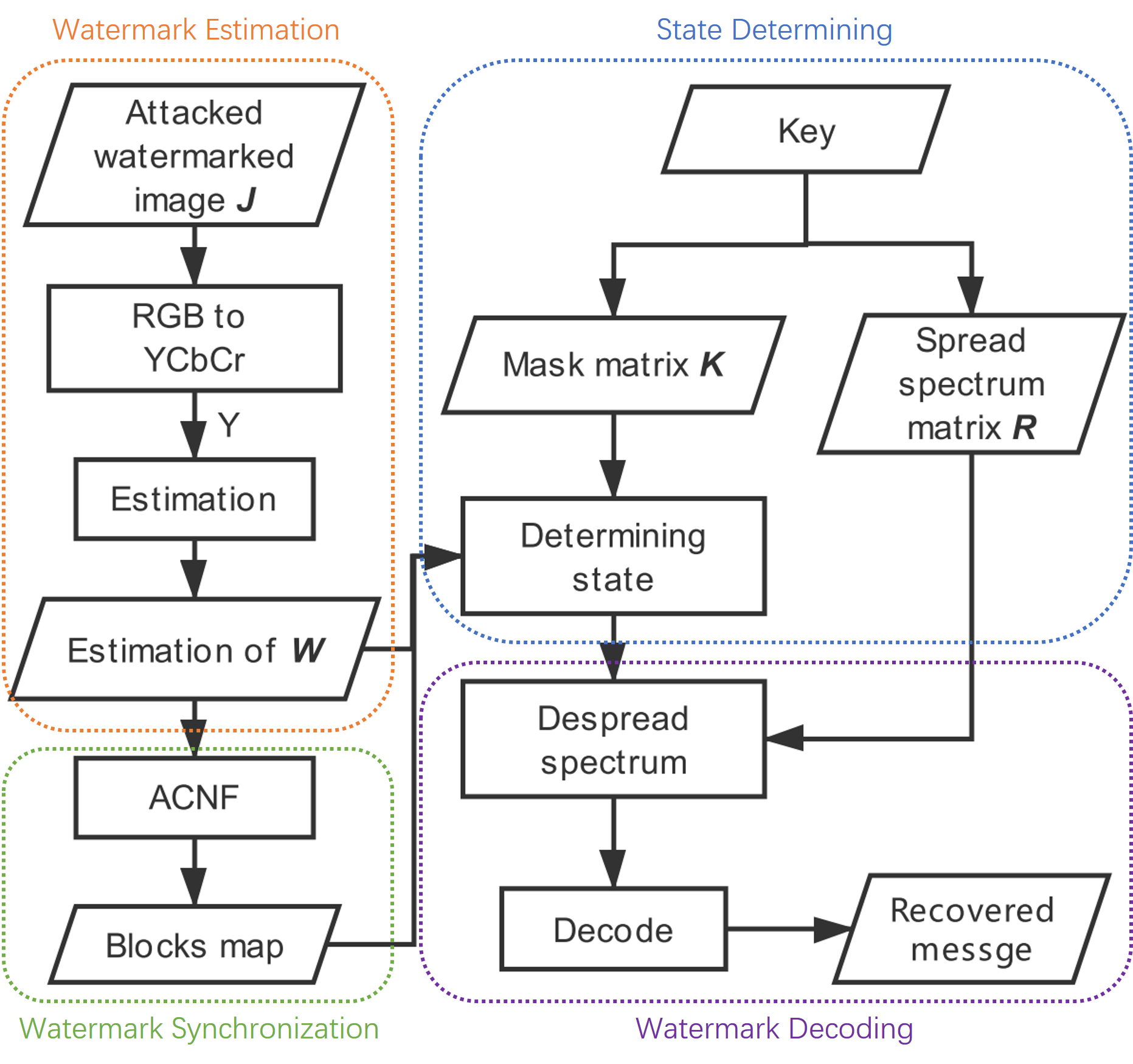}
  \caption{Watermark extraction process.}\label{extraction}
  \end{center}
\end{figure}

As a blind watermarking scheme, the watermark extractor has no prior knowledge of the host image. However, the key is shared between encoder and decoder, so the extractor can generate the same spread-spectrum matrix $\boldsymbol{R}$ and the mask matrix $\boldsymbol{K}$. The watermark extraction process is illustrated in Fig.~\ref{extraction} and can be divided into four main modules: watermark estimation, watermark synchronization, watermark state determining, and watermark decoding. These modules are further explained as follows.

\subsection{Watermark Estimation}
Part of the discussion in this section refers to \cite{book_wiener}. The watermark component is predicted from the attacked watermarked image $\boldsymbol{J}$ received by the extractor. We denote the distortion and noise as $\boldsymbol{n}$. Then $\boldsymbol{J}$ can be represented as:
\begin{equation}\label{J_rp}
    J(i,j)=I(i,j)+W(i,j)+n(i,j).
\end{equation}
$\boldsymbol{n}$ is assumed to be zero-mean white noise for the purpose of tractability. Considering $\boldsymbol{w}_m$ is almost a random matrix and $\boldsymbol{W}$ is generated by $\boldsymbol{w}_m$, $\boldsymbol{W}$ is similar to an additive random noise just like $\boldsymbol{n}$. So $\boldsymbol{J}$ and $\boldsymbol{I}$ have the same local mean $\boldsymbol{\mu}$. Eq.~(\ref{J_rp}) can be further represented as: 
\begin{equation}\label{J_de}
    J^{\prime}(i,j)=I^{\prime}(i,j)+W(i,j)+n(i,j)
\end{equation}
where $\boldsymbol{I}^{\prime}$ and $\boldsymbol{J}^{\prime}$ are residual components such that $\boldsymbol{I}^{\prime}=\boldsymbol{I}-\boldsymbol{\mu}$,  $\boldsymbol{J}^{\prime}=\boldsymbol{J}-\boldsymbol{\mu}$.
We hope to find a kind of $\boldsymbol{h}$ so that $\boldsymbol{W}$ can be estimated as follows:
\begin{equation}\label{h_def}
    \widehat{\boldsymbol{W}}=\boldsymbol{J}^{\prime}\otimes\boldsymbol{h}
\end{equation}
where $\otimes$ represents the convolution operation and $\widehat{\boldsymbol{W}}$ is an estimate of $\boldsymbol{W}$. $\boldsymbol{h}$ is the convolution kernel which can minimize the mean-square error:
\begin{equation}
    \boldsymbol{h}= \operatorname{arg\,min}\, E\{(\boldsymbol{W}-\widehat{\boldsymbol{W}})^{2}\}
\end{equation}
where $E\{\cdot \}$ represents the expectation. Substituting $\widehat{\boldsymbol{W}}$ with Eq.~(\ref{h_def}), the mean square error can be rewritten as:
\begin{equation}\label{MSE}
    E\{(\boldsymbol{W}-\widehat{\boldsymbol{W}})^{2}\}=E\{(\boldsymbol{W}-\boldsymbol{J}^{\prime}\otimes\boldsymbol{h}\}.
\end{equation}
Substituting $\boldsymbol{J}^{\prime}$ with Eq.~(\ref{J_de}), Eq.~(\ref{MSE}) can be rewritten in frequency domain as:
\begin{equation}\label{W_ex}
    \begin{aligned}
        E\{(\boldsymbol{W}-\widehat{\boldsymbol{W}})^{2}\}&=E\{(\overline{\boldsymbol{W}}-\overline{\boldsymbol{J}^{\prime}}\boldsymbol{H})^{2}\}\\
        &=E\{(\overline{\boldsymbol{W}}-(\overline{\boldsymbol{I}^{\prime}}+\overline{\boldsymbol{W}}+\boldsymbol{N})\boldsymbol{H})^{2}\}\\
        &=E\{(\overline{\boldsymbol{W}}(1-\boldsymbol{H})-\overline{\boldsymbol{I}^{\prime}}\boldsymbol{H}-\boldsymbol{N}\boldsymbol{H})^{2}\}\\
        &=E\{\overline{\boldsymbol{W}}^{2}(1-\boldsymbol{H})^{2}+\overline{\boldsymbol{I}^{\prime}}^{2}\boldsymbol{H}^{2}+\boldsymbol{N}^{2}\boldsymbol{H}^{2}+\boldsymbol{Z}\}
    \end{aligned}
\end{equation}
where $\overline{\boldsymbol{I}^{\prime}}$, $\overline{\boldsymbol{J}^{\prime}}$, $\overline{\boldsymbol{W}}$, $\boldsymbol{N}$, $\boldsymbol{H}$ are the frequency form of $\boldsymbol{I}^{\prime}$, $\boldsymbol{J}^{\prime}$, $\boldsymbol{W}$, $\boldsymbol{n}$, $\boldsymbol{h}$, respectively. $\boldsymbol{Z}$ is the summation of  the cross product. Specifically,
\begin{equation}
    \begin{split}
        \boldsymbol{Z}&=-\overline{\boldsymbol{W}}(1-\boldsymbol{H})\overline{\boldsymbol{I}^{\prime}}^{*}\boldsymbol{H}^{*}-\overline{\boldsymbol{W}}^{*}(1-\boldsymbol{H})^{*}\overline{\boldsymbol{I}^{\prime}}\boldsymbol{H}-\overline{\boldsymbol{W}}(1-\boldsymbol{H})\boldsymbol{N}^{*}\boldsymbol{H}^{*}\\
        &\quad-\overline{\boldsymbol{W}}^{*}(1-\boldsymbol{H})^{*}\boldsymbol{N}\boldsymbol{H}+\overline{\boldsymbol{I}^{\prime}}\boldsymbol{H}\boldsymbol{N}^{*}\boldsymbol{H}^{*}+\overline{\boldsymbol{I}^{\prime}}^{*}\boldsymbol{H}^{*}\boldsymbol{N}\boldsymbol{H}\\
        &=-\overline{\boldsymbol{W}}\overline{\boldsymbol{I}^{\prime}}^{*}(1-\boldsymbol{H})\boldsymbol{H}^{*}-\overline{\boldsymbol{W}}^{*}\overline{\boldsymbol{I}^{\prime}}(1-\boldsymbol{H})^{*}\boldsymbol{H}-\overline{\boldsymbol{W}}\boldsymbol{N}^{*}(1-\boldsymbol{H})\boldsymbol{H}^{*}\\
        &\quad-\overline{\boldsymbol{W}}^{*}\boldsymbol{N}(1-\boldsymbol{H})^{*}\boldsymbol{H}+\overline{\boldsymbol{I}^{\prime}}\boldsymbol{N}^{*}\boldsymbol{H}\boldsymbol{H}^{*}+\overline{\boldsymbol{I}^{\prime}}^{*}\boldsymbol{N}\boldsymbol{H}^{*}\boldsymbol{H}.
    \end{split}
\end{equation}

Note that the cross-correlation of two variables can be calculated by the conjugate product of their frequency form. Because the watermark $\boldsymbol{W}$, residual component $\boldsymbol{I}^{\prime}$ and noise $\boldsymbol{n}$ are independent, the expectation of their cross-correlation should be zero. So the expectation of $\boldsymbol{Z}$ is
\begin{equation}\label{O_ex}
    \begin{aligned}
        E\{\boldsymbol{Z}\}&=-E\{\overline{\boldsymbol{W}}\overline{\boldsymbol{I}^{\prime}}^{*}\}(1-\boldsymbol{H})\boldsymbol{H}^{*}-E\{\overline{\boldsymbol{W}}^{*}\overline{\boldsymbol{I}^{\prime}}\}(1-\boldsymbol{H})^{*}\boldsymbol{H}\\
        &\quad-E\{\overline{\boldsymbol{W}}\boldsymbol{N}^{*}\}(1-\boldsymbol{H})\boldsymbol{H}^{*}-E\{\overline{\boldsymbol{W}}^{*}\boldsymbol{N}\}(1-\boldsymbol{H})^{*}\boldsymbol{H}\\
        &\quad+E\{\overline{\boldsymbol{I}^{\prime}}\boldsymbol{N}^{*}\}\boldsymbol{H}\boldsymbol{H}^{*}+E\{\overline{\boldsymbol{I}^{\prime}}^{*}\boldsymbol{N}\}\boldsymbol{H}^{*}\boldsymbol{H}\\
        &=0.
    \end{aligned}
\end{equation}
Combining Eq.~(\ref{W_ex}) and Eq.~(\ref{O_ex}), we have:
\begin{equation}\label{H_qf}
    \begin{aligned}
    E\{(\boldsymbol{W}-\widehat{\boldsymbol{W}})^{2}\}&=E\{\overline{\boldsymbol{W}}^{2}\}(1-\boldsymbol{H})^{2}+E\{\overline{\boldsymbol{I}^{\prime}}^{2}\}\boldsymbol{H}^{2}\\
    &\quad+E\{\boldsymbol{N}^{2}\}\boldsymbol{H}^{2}+E\{\boldsymbol{Z}\}\\
    &=P_{\boldsymbol{W}}(1-\boldsymbol{H})^{2}+P_{\boldsymbol{I}^{\prime}}\boldsymbol{H}^{2}+P_{\boldsymbol{N}}\boldsymbol{H}^{2}+0\\
    &=P_{\boldsymbol{W}}(1-\boldsymbol{H})^{2}+(P_{\boldsymbol{I}^{\prime}}+P_{\boldsymbol{N}})\boldsymbol{H}^{2}
    \end{aligned}
\end{equation}
where $P_{\boldsymbol{W}}$, $P_{\boldsymbol{I}^{\prime}}$, $P_{\boldsymbol{N}}$ are the power spectrum of $\boldsymbol{W}$, $\boldsymbol{I}^{\prime}$, $\boldsymbol{N}$. We find that the mean-square error is a quadratic function about $\boldsymbol{H}$. To find the minimum error value, the derivative of Eq.~(\ref{H_qf}) is calculated and set to zero. Then we have:
\begin{equation}\label{H_value}
    \begin{aligned}
        \boldsymbol{H}&=\frac{P_{\boldsymbol{W}}}{P_{\boldsymbol{W}}+P_{\boldsymbol{I}^{\prime}}+P_{\boldsymbol{N}}}\\
        &=\frac{P_{\boldsymbol{W}}}{P_{\boldsymbol{J}^{\prime}}}
    \end{aligned}
\end{equation}
where $P_{\boldsymbol{J}^{\prime}}$ is the power spectrum of $\boldsymbol{J}^{\prime}$. From Eq.~(\ref{H_value}), we can get that $h(i,j)$ is a scaled impulse given by:
\begin{equation}\label{h_impulse}
    h(i,j)=\frac{P_{W}}{P_{J^{\prime}}}\delta(i,j)
\end{equation}
where $\delta$ is a unit impulse. Substituting the right side of Eq.~(\ref{h_def}) with Eq.~(\ref{h_impulse}), the watermark $\boldsymbol{W}$ within the local region can be expressed as:
\begin{equation}
    \begin{aligned}
        \widehat{\boldsymbol{W}}&=\boldsymbol{J}^{\prime}\otimes\frac{P_{\boldsymbol{W}}}{P_{\boldsymbol{J}^{\prime}}}\boldsymbol{\delta}\\
        &=(\boldsymbol{J}-\boldsymbol{\mu})\frac{P_{\boldsymbol{W}}}{P_{\boldsymbol{J}^{\prime}}}.
    \end{aligned}
\end{equation}
Considering $\boldsymbol{J}^{\prime}$ and $W$ are both zero mean in the local region, their power spectrum are their local variance. So the mean-square error minimized estimation of $\boldsymbol{W}$ can be calculated by:
\begin{equation}
    \widehat{\boldsymbol{W}}=(\boldsymbol{J}-\boldsymbol{\mu})\frac{\boldsymbol{\sigma}^{2}_{\boldsymbol{W}}}{\boldsymbol{\sigma}^{2}_{\boldsymbol{J}^{\prime}}}.
\end{equation}
where $\boldsymbol{\sigma}^{2}_{\boldsymbol{W}}$, $\boldsymbol{\sigma}^{2}_{\boldsymbol{J}^{\prime}}$ are the local variance of $\boldsymbol{W}$ and $\boldsymbol{J}^{\prime}$. We can calculate $\boldsymbol{\sigma}^{2}_{\boldsymbol{J}^{\prime}}$ from the attacked watermarked image $\boldsymbol{J}$. Note that $\boldsymbol{W}$ is similar to a random noise so its local variance $\boldsymbol{\sigma}^{2}_{\boldsymbol{W}}$ depends on its embedding strength $\boldsymbol{s}$, which can be estimated from $\boldsymbol{J}$ using Eq.~(\ref{adaptive strength}).

\subsection{Watermark Synchronization based on Symmetry}
To synchronize the masked watermark unit $\boldsymbol{w}_m$, the symmetry of $\widehat{\boldsymbol{W}}$ should be calculated first. In Section.~\ref{illustration_of_SP}, We have defined the symmetry of symmetrical watermark and proposed Eq.~(\ref{S1_def}) to calculate it. Also in Section.~\ref{illustration_of_SP}, we believe that a lower computational cost version of Eq.~(\ref{S1_def}) is needed.

First of all, the Eq.~(\ref{S1_def}) can be rewritten as:
\begin{equation}\label{S_def}
    \begin{aligned}
        S(i,j)&=\sum_{x} \sum_{y} \widehat{W}(i-x,j-y)\widehat{W}(i+x,j+y)\\
        &=\sum_{x} \sum_{y} \widehat{W}(x,y)\widehat{W}(2i-x,2j-y)
    \end{aligned}
\end{equation}
where $x$ and $y$ run over all values that lead to legal subscripts of $\widehat{\boldsymbol{W}}$.
Defining a temporary matrix $\boldsymbol{T}$, $T(2i,2j)=S(i,j)$, we have:
\begin{equation}\label{Temp_def}
    \begin{aligned}
        T(2i,2j)&=\sum_{x} \sum_{y} \widehat{W}(x,y)\widehat{W}(2i-x,2j-y)\\
        T(u,v)&=\sum_{x} \sum_{y} \widehat{W}_p(x,y)\widehat{W}_p(u-x,v-y)
    \end{aligned}
\end{equation}
where $\widehat{\boldsymbol{W}}_p$ is $\widehat{\boldsymbol{W}}$ zero-padding to doubly the original size.
We discover that $\boldsymbol{T}$ is the auto-convolution of $\widehat{\boldsymbol{W}}_p$. Just like periodicity and auto-correlation function (ACF), symmetry has a relationship with auto-convolution function. In this paper, \textbf{A}uto-\textbf{C}o\textbf{N}volution \textbf{F}unction is abbreviated as ACNF to distinguish it from ACF. So the convolution theorem could be used to get the frequency form of Eq.~(\ref{Temp_def}) as follows:
\begin{equation}
    \boldsymbol{T}=IFFT[FFT(\widehat{\boldsymbol{W}}_p)FFT(\widehat{\boldsymbol{W}}_p)]
\end{equation}
where $FFT$ represents the fast Fourier transform and $IFFT$ is the corresponding inverse transform. According to the definition of $\boldsymbol{T}$, $\boldsymbol{T}$ is actually the doubly upsampling of $\boldsymbol{S}$, so the symmetry $\boldsymbol{S}$ of $\widehat{\boldsymbol{W}}$ can be calculated by:
\begin{equation}\label{S_freq}
    \boldsymbol{S}=\mathcal{D}(IFFT[FFT(\widehat{\boldsymbol{W}}_p)FFT(\widehat{\boldsymbol{W}}_p)])
\end{equation}
where $\mathcal{D}(\cdot)$ is a downsampling function which scales its input matrix to the half size. Using the frequency form of ACNF, from Eq.~(\ref{S_def}) to Eq.~(\ref{S_freq}), the computational cost of calculating $\boldsymbol{S}$ is greatly reduced.

\begin{figure}[t]
    \begin{center}
    \subfloat[Original symmetrical peaks.]{
        {\centering\includegraphics[width=1.6in]{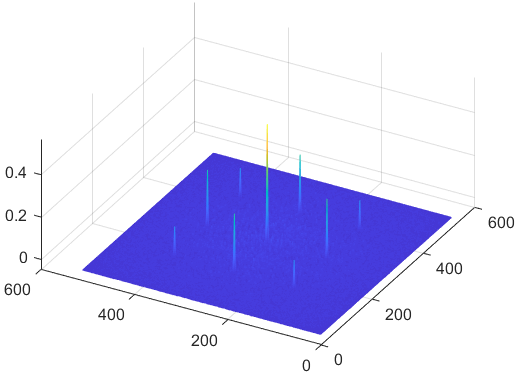}
        }
    }
    \subfloat[Uniform symmetrical peaks.]{
        {\centering\includegraphics[width=1.6in]{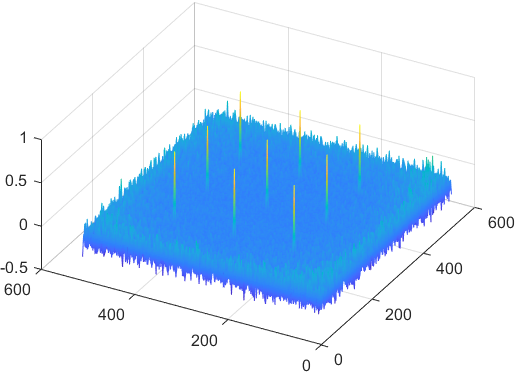}
        }
    }
    \end{center}
\caption{Original and uniform symmetrical peaks generated from the same symmetrical watermark. The symmetrical watermark has $4\times4$ watermark units.}
\label{uniform}
\end{figure}

Another problem is that, in Eq.~(\ref{S_def}), different $i$, $j$ will contribute different summation size. As a result, $\boldsymbol{S}$ has higher peaks in the middle region, as shown in Fig.~\ref{uniform}(a). This phenomenon is also reflected in the brightness of symmetrical peaks in Fig.~\ref{samples_of_SP}. Actually, in the spatial domain, the uniform $\boldsymbol{S}^{\prime}$ is easy to get by dividing $\boldsymbol{S}$ by its summation size as follows:
\begin{equation}\label{S_unit}
    S^{\prime}(i,j)=\frac{\sum_{x}\sum_{y}\widehat{W}(i+x,j+y)\widehat{W}(i-x,j-y)}{\sum_{x}\sum_{y}1}
\end{equation}
where $S^{\prime}(i,j)$ is a uniform symmetry generated by scaling $S(i,j)$ to the unit value. It is hard to find the frequency form of Eq.~(\ref{S_unit}) directly. Kang \emph{et.al} \cite{ULPM} propose customized phase correlation to solve a similar problem and get an approximate solution. In our scheme, we first calculate the numerator and denominator of Eq.~(\ref{S_unit}) separately and then calculate their quotient. We note that the denominator also has frequency form just like the numerator. So the uniform symmetry $\boldsymbol{S}^{\prime}$ can be calculated by:
\begin{equation}
    \boldsymbol{S}^{\prime}=\mathcal{D}\left(\frac{IFFT[FFT(\widehat{\boldsymbol{W}}_p)FFT(\widehat{\boldsymbol{W}}_p)]}{IFFT[FFT(\boldsymbol{O}_p)FFT(\boldsymbol{O}_p)]}\right)
\end{equation}
where $\boldsymbol{O}_p$ is the zero-padding of $\boldsymbol{O}$ and $\boldsymbol{O}$ is a matrix of ones having the same size with $\widehat{\boldsymbol{W}}$. Fig.~\ref{uniform} shows the symmetrical peaks of original $\boldsymbol{S}$ and uniform $\boldsymbol{S}^{\prime}$. The peaks of $\boldsymbol{S}^{\prime}$ have similar height.

\begin{figure}[t]
    \begin{center}
    \subfloat[Watermarked image rotated by \newline $15^\circ$, translated by (-16,-16), and \newline cropped to 75\%.]{
        {\centering\includegraphics[width=1.55in]{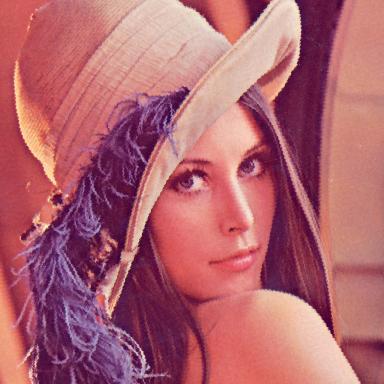}
        }
    }
    \subfloat[$\boldsymbol{M}$ corresponding to (a).]{
        {\centering\includegraphics[width=1.55in]{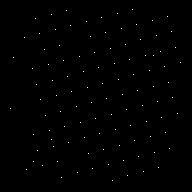}
        }
    }\\
    \subfloat[Watermarked image attacked by \newline RBA.]{
        {\centering\includegraphics[width=1.55in]{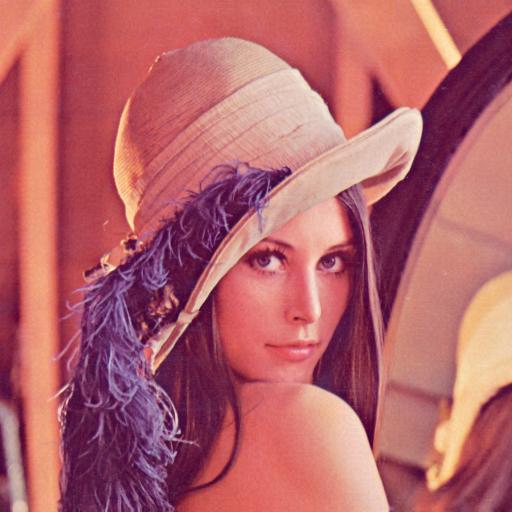}
        }
    }
    \subfloat[$\boldsymbol{M}$ corresponding to (c).]{
        {\centering\includegraphics[width=1.55in]{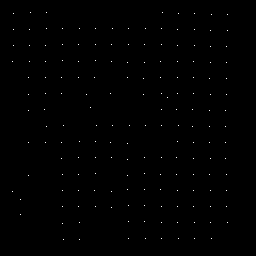}
        }
    }
    \end{center}
\caption{Watermarked images under geometric distortions and their corresponding watermark units map $\boldsymbol{M}$. The embedded symmetrical watermark has $16\times16$ watermark units.}
\label{RST_map}
\end{figure}

To separate symmetrical peaks from noise peaks and get watermark units map $\boldsymbol{M}$, an adaptive threshold is applied as follows:
\begin{equation}
    \boldsymbol{M}=\left\{\begin{array}{ll}{1} & {, \quad if\quad\boldsymbol{S}^{\prime}>\boldsymbol{\mu}_{\boldsymbol{S}^{\prime}}+\beta\boldsymbol{\sigma}^{2}_{\boldsymbol{S}^{\prime}}} \\ {0} & {, \quad otherwise}\end{array}\right.
\end{equation}
where $\boldsymbol{\mu}_{\boldsymbol{S}^{\prime}}$ and $\boldsymbol{\sigma}^{2}_{\boldsymbol{S}^{\prime}}$ denote the local average and the standard deviation of $\boldsymbol{S}^{\prime}$, respectively. And $\beta$ is an empirical coefficient which is set from 3.0 to 4.3. $\boldsymbol{M}$ is a binary matrix in which element 1's represent the possible corner points of watermark units and we can use it to resist various geometric distortions. As shown in Fig.~\ref{RST_map}, $\boldsymbol{M}$ clearly shows the position of the watermark units under geometric distortions.

\begin{figure}[t]
    \begin{minipage}{0.65in}
      \centerline{\includegraphics[width=1\textwidth]{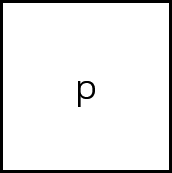}}
      \caption*{\emph{state 1}}
    \end{minipage}
    \hfill
    \begin{minipage}{0.65in}
      \centerline{\includegraphics[width=1\textwidth]{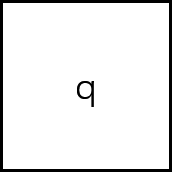}}
      \caption*{\emph{state 2}}
    \end{minipage}
    \hfill
    \hspace{.1in}
    \begin{minipage}{0.65in}
      \centerline{\includegraphics[width=1\textwidth]{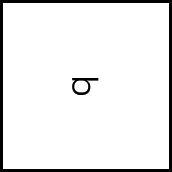}}
      \caption*{\emph{state 5}}
    \end{minipage}
    \hfill
    \begin{minipage}{0.65in}
      \centerline{\includegraphics[width=1\textwidth]{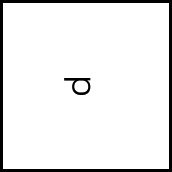}}
      \caption*{\emph{state 6}}
    \end{minipage}
    \vfill
    \vspace{.1in}
    \begin{minipage}{0.65in}
      \centerline{\includegraphics[width=1\textwidth]{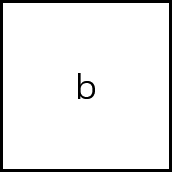}}
      \caption*{\emph{state 3}}
    \end{minipage}
    \hfill
    \begin{minipage}{0.65in}
      \centerline{\includegraphics[width=1\textwidth]{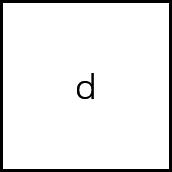}}
      \caption*{\emph{state 4}}
    \end{minipage}
    \hfill
    \hspace{.1in}
    \begin{minipage}{0.65in}
      \centerline{\includegraphics[width=1\textwidth]{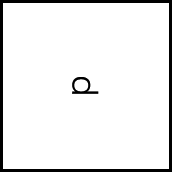}}
      \caption*{\emph{state 7}}
    \end{minipage}
    \hfill
    \begin{minipage}{0.65in}
      \centerline{\includegraphics[width=1\textwidth]{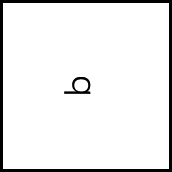}}
      \caption*{\emph{state 8}}
    \end{minipage}
    \caption{Four original states and four extra states generated by rotation.}
    \label{pbqd}
\end{figure}

\subsection{Watermark State Determining}\label{state_determined}
Before despread spectrum and decoding the masked watermark unit $\boldsymbol{w}_{m}$, the state of $\boldsymbol{w}_{m}$ should be determined. Flipping offers the watermark symmetry but introduces additional states. As Fig.~\ref{pbqd} shows, the original watermark unit has four states. And if rotation is considered, the states increase to eight. Rather than introducing a reference watermark in the watermark unit, we propose a state determining method by combining mask matrix $\boldsymbol{K}$ and the statistical characteristics of $\boldsymbol{w}_{m}$.

The state of $\boldsymbol{w}_{m}$ without any flipping and rotation is denoted as \emph{state 1} and the rest are denoted as \emph{state 2} to \emph{8}. Firstly, according to the watermark units map $\boldsymbol{M}$, we select a watermark unit and restore its geometric distortions to get $\boldsymbol{w}_s$, whose state is unknown. Before de-masking $\boldsymbol{w}_s$ with $\boldsymbol{K}$, $\boldsymbol{w}_s$ should be flipped and rotated from current state to \emph{state 1}. Set up a null hypothesis as follows:
\begin{center}
    $H_{0}$: $\boldsymbol{w}_{s}$ is \textbf{not} \emph{state 1}. 
\end{center}
Contrary to $H_{0}$, $\boldsymbol{w}_s$ is de-masked with $\boldsymbol{K}$ directly to get $\boldsymbol{w}_r$. According to $H_{0}$, $\boldsymbol{w}_s$ is wrong de-masked so $\boldsymbol{w}_r$ is almost a random matrix and has no property of spread-spectrum matrix $\boldsymbol{R}$. We divide $\boldsymbol{w}_r$ into non-overlapped small blocks which have the same size with $\boldsymbol{R}$, and use $\boldsymbol{w}_r^{i,j}$ to represent the small block $i$-th in row and $j$-th in column. Then $\boldsymbol{w}_r^{i,j}$ is normalized to the mean value of 0 and the variance of 1. It should be noted that $\boldsymbol{R}$ is also zero-mean and one-variance. Matrix $\boldsymbol{x}$ is defined by:
\begin{equation}
    x(p,q)=w_r^{i,j}(p,q)R(p,q)
\end{equation}
where $p$, $q$ are integer ranging from 1 to $L_{R}$. According to $H_{0}$, $w_r^{i,j}(p,q)$ is independent to $R(p,q)$. So the expectation and the variance of $x(p,q)$ are:
\begin{equation}
    \begin{aligned}
        \mu_{x}&=E\{x(p,q)\}\\
        &=E\{w_r^{i,j}(p,q)\}E\{R(p,q)\}\\
        &=0
    \end{aligned}
\end{equation}
\begin{equation}
    \begin{aligned}
        \sigma_{x}^{2}&=D\{x(p,q)\}\\
        &=E\{(w_r^{i,j}(p,q)R(p,q))^{2}\}-E\{w_r^{i,j}(p,q)R(p,q)\}^{2}\\
        &=D\{w_r^{i,j}(p,q)\}D\{R(p,q)\}-E\{w_r^{i,j}(p,q)\}^{2}E\{R(p,q)\}^{2}\\
        &=1.
    \end{aligned}
\end{equation}
Consider the following equation:
\begin{equation}
    \begin{aligned}
        y&=\frac{1}{\sqrt{L_{R}^{2}}\sigma_{x}}\sum_{p}^{L_{R}}\sum_{q}^{L_{R}}(x(p,q)-\mu_{x})\\
        &=\frac{1}{L_{R}}\sum_{p}^{L_{R}}\sum_{q}^{L_{R}}x(p,q).
    \end{aligned}
\end{equation}
If the size of $\boldsymbol{R}$ is sufficiently large, $y$ will be a random variable following a standard normal distribution $N(0,1)$ by exploiting the central limit theorem. One selected watermark unit $\boldsymbol{w}_{s}$ can offer a few samples and we can roughly evaluate whether $y$ follows $N(0,1)$ according to these samples.

\begin{figure}[t]
    \begin{center}
    \subfloat[\textbf{not} \emph{state 1}]{
        {\centering\includegraphics[width=.75in]{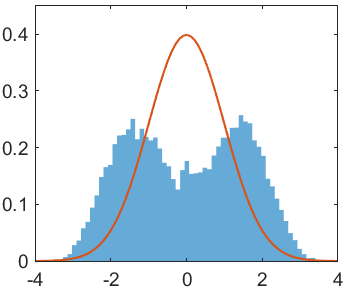}
        }
    }
    \subfloat[\textbf{not} \emph{state 2}]{
        {\centering\includegraphics[width=.75in]{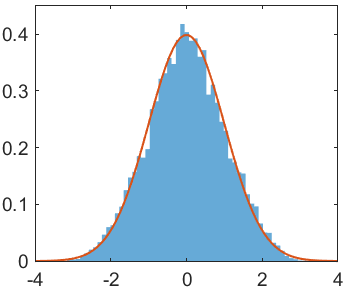}
        }
    }
    \subfloat[\textbf{not} \emph{state 3}]{
        {\centering\includegraphics[width=.75in]{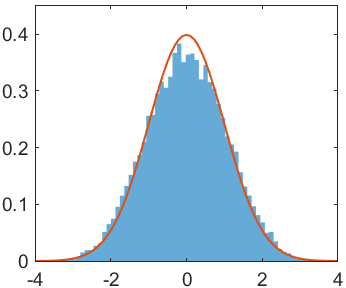}
        }
    }
    \subfloat[\textbf{not} \emph{state 4}]{
        {\centering\includegraphics[width=.75in]{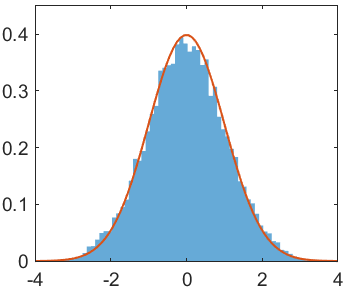}
        }
    }\\
    \subfloat[\textbf{not} \emph{state 5}]{
        {\centering\includegraphics[width=.75in]{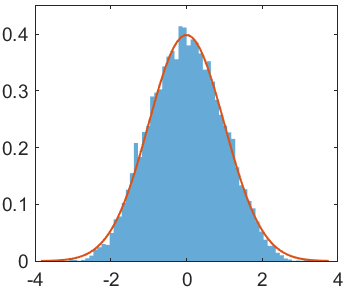}
        }
    }
    \subfloat[\textbf{not} \emph{state 6}]{
        {\centering\includegraphics[width=.75in]{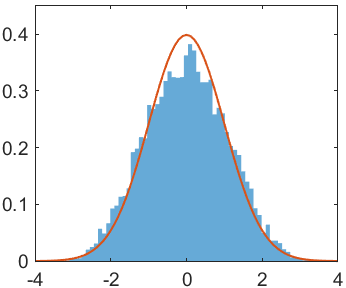}
        }
    }
    \subfloat[\textbf{not} \emph{state 7}]{
        {\centering\includegraphics[width=.75in]{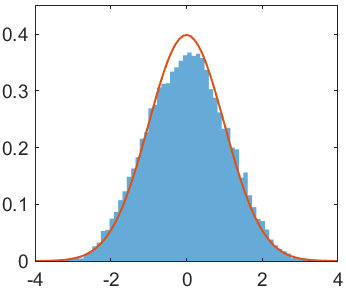}
        }
    }
    \subfloat[\textbf{not} \emph{state 8}]{
        {\centering\includegraphics[width=.75in]{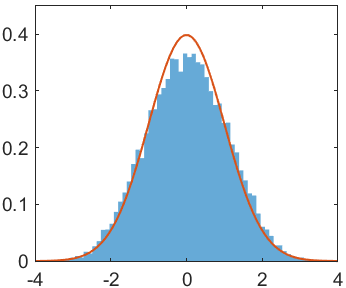}
        }
    }
    \end{center}
    
    \caption{Different distribution of $y$ with different null hypotheses of $\boldsymbol{w}_s$ when $\boldsymbol{w}_s$ is \emph{state 1} actually. The red line follows $N(0,1)$.}
    \label{CLT}
\end{figure}

Note that $H_{0}$ actually makes hypothesis to all blocks' state besides the state of $\boldsymbol{w}_{s}$ because the states of watermark units are related. For example, according to the \emph{flipping rule} mentioned in Section~\ref{W_gen}, a watermark unit $\boldsymbol{w}$ on right side of $\boldsymbol{w}_{s}$ is generated by flipping $\boldsymbol{w}_{s}$ horizontally. So if $\boldsymbol{w}_{s}$ is \emph{state 1}, the $\boldsymbol{w}$ on right side of $\boldsymbol{w}_{s}$ will be \emph{state 2}. And if $\boldsymbol{w}_{s}$ is not \emph{state 1}, the $\boldsymbol{w}$ on right side of $\boldsymbol{w}_{s}$ will not be \emph{state 2}. Similarly, combining $H_{0}$ and flipping rule, it is easy to conclude an equivalent null hypothesis $H_{0}^{\prime}$:
\begin{center}
    $H_{0}^{\prime}$: $\boldsymbol{w}_{s}$ is \textbf{not} \emph{state 1};\\
    $\boldsymbol{w}$ on right side of $\boldsymbol{w}_{s}$ is \textbf{not} \emph{state 2};\\
    $\boldsymbol{w}$ on upper side of $\boldsymbol{w}_{s}$ is \textbf{not} \emph{state 3};\\
    \ldots\ .
\end{center}
$H_{0}^{\prime}$ makes hypothesis to all watermark units. To every block, we can get a set of samples of $y$. So we have enough samples of $y$ from these blocks and could evaluate the distribution of $y$ accurately. Fig.~\ref{CLT}(a) shows the distribution of $y$ with $H_{0}^{\prime}$ when the selected $\boldsymbol{w}_{s}$ is \emph{state 1}. It is obvious that $y$ deviates on a large scale from $N(0,1)$, so $H_{0}^{\prime}$ is rejected. And $H_{0}$ is also rejected since $H_{0}$ and $H_{0}^{\prime}$ are equivalent. Finally, we can get the conclusion that $\boldsymbol{w}_{s}$ is \emph{state 1}.

Meanwhile, as Fig.~\ref{CLT} shows, if we make a different null hypothesis, $y$ will follow $N(0,1)$. In practice, we use Kullback-Leibler divergence (KLD) to measure the distance between $y$ and $N(0,1)$. To eight different null hypotheses, the distance between $y$ and $N(0,1)$ will be calculated respectively. Then the hypothesis corresponding to the farthest distance is rejected, and the state of $\boldsymbol{w}_{s}$ will be determined.

\subsection{Watermark Decoding}
Now we get the states of all watermark units and restore all available watermark units to \emph{state 1}. All these available blocks are accumulated to get $\widehat{\boldsymbol{w}}$. Similarly, $\widehat{\boldsymbol{w}}$ is divided to non-overlapped small blocks which have the same size with spread-spectrum matrix $\boldsymbol{R}$ and they are denoted as $\widehat{\boldsymbol{w}}^{i,j}$. To despread $\widehat{\boldsymbol{w}}$, the correlation value $\rho_{i,j}$ between $\widehat{\boldsymbol{w}}^{i,j}$ and $\boldsymbol{R}$ is obtained as:
\begin{equation}
    \rho_{i,j}=\sum_{p}^{L_{R}}\sum_{q}^{L_{R}}\widehat{w}^{i,j}(p,q)R(p,q)
\end{equation}
Then the message bit is determined by:
\begin{equation}
    \widehat{m}_{i,j}=\left\{\begin{array}{ll}{0} & {, \quad \rho_{i,j}<0} \\ {1} & {, \quad \rho_{i,j} \geq 0}\end{array}\right.
\end{equation}
where $\widehat{m}_{i,j}$ is the extracted message bit. The obtained message matrix $\widehat{\boldsymbol{m}}$ is now reshaped and ECC decoded to recover the embedding message $\boldsymbol{m}^{\prime}$.

\begin{figure*}[t]
    \begin{minipage}{0.8in}
      \centerline{\includegraphics[width=1\textwidth]{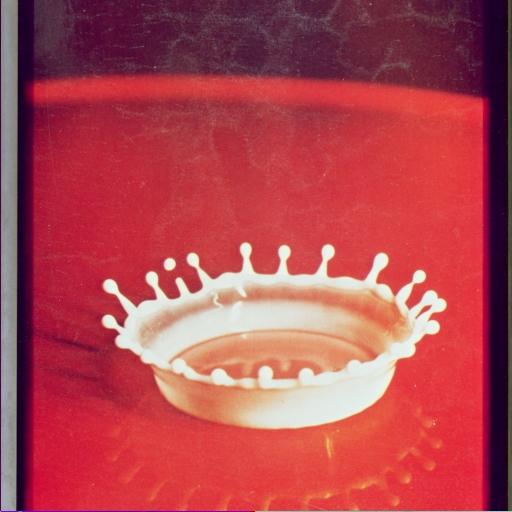}}
    \end{minipage}
    \hfill
    \begin{minipage}{0.8in}
      \centerline{\includegraphics[width=1\textwidth]{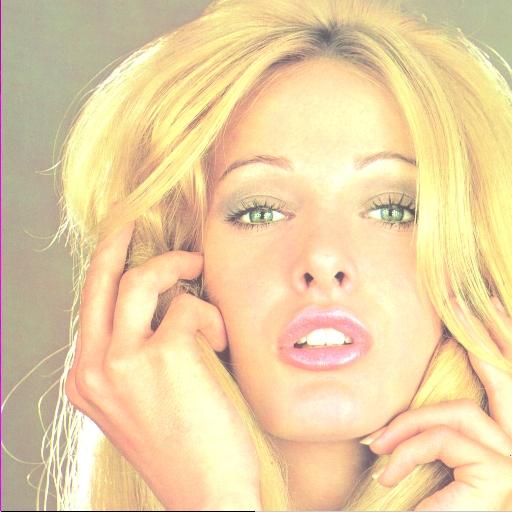}}
    \end{minipage}
    \hfill
    \begin{minipage}{0.8in}
      \centerline{\includegraphics[width=1\textwidth]{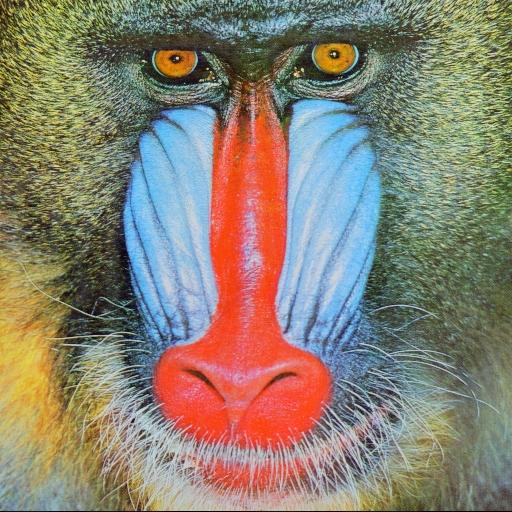}}
    \end{minipage}
    \hfill
    \begin{minipage}{0.8in}
      \centerline{\includegraphics[width=1\textwidth]{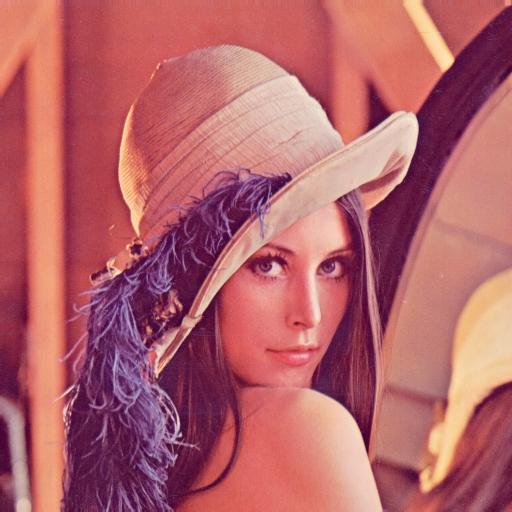}}
    \end{minipage}
    \hfill
    \begin{minipage}{0.8in}
      \centerline{\includegraphics[width=1\textwidth]{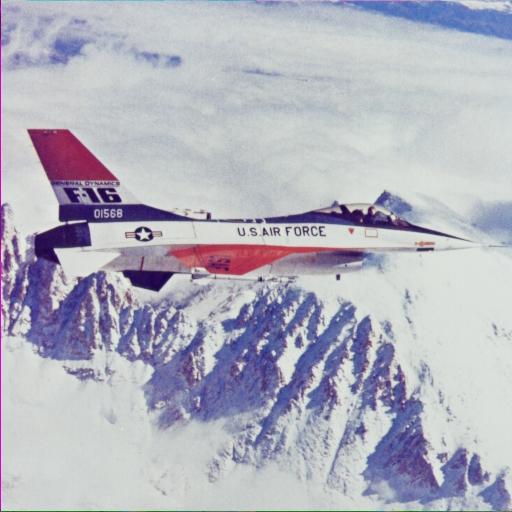}}
    \end{minipage}
    \hfill
    \begin{minipage}{0.8in}
      \centerline{\includegraphics[width=1\textwidth]{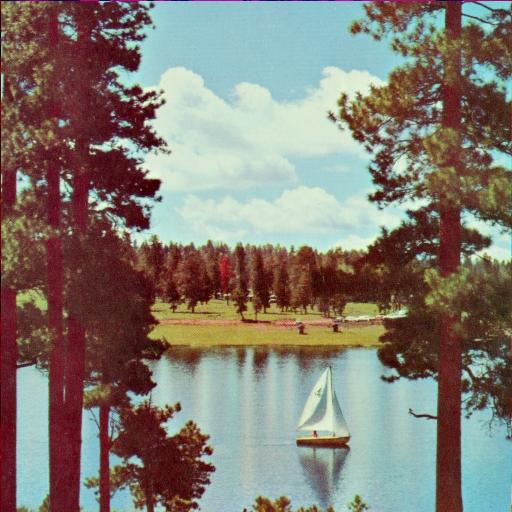}}
    \end{minipage}
    \hfill
    \begin{minipage}{0.8in}
      \centerline{\includegraphics[width=1\textwidth]{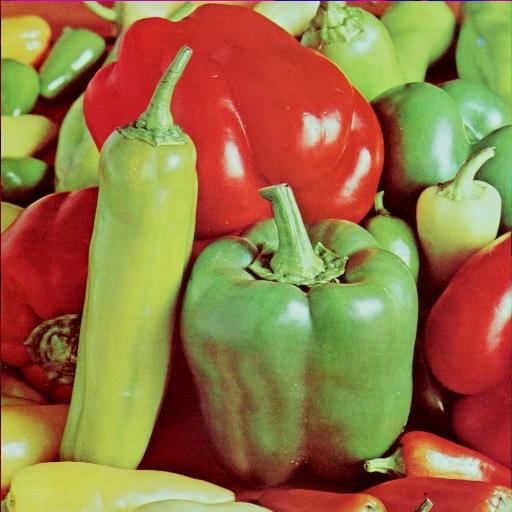}}
    \end{minipage}
    \hfill
    \begin{minipage}{0.8in}
      \centerline{\includegraphics[width=1\textwidth]{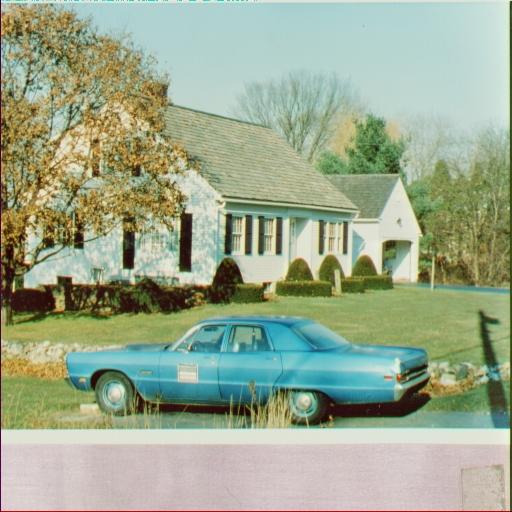}}
    \end{minipage}
    \vfill
    \vspace{.1in}
    \begin{minipage}{0.8in}
      \centerline{\includegraphics[width=1\textwidth]{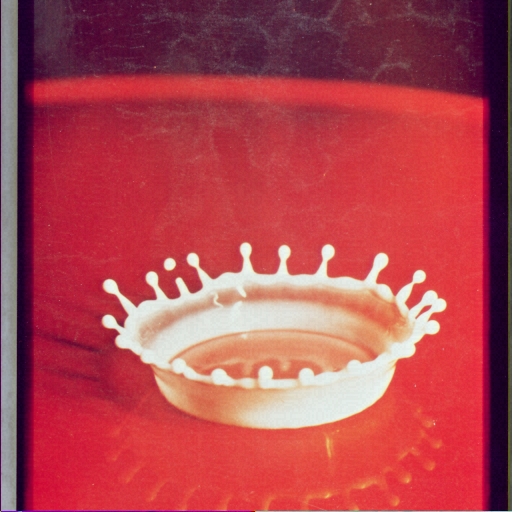}}
    \end{minipage}
    \hfill
    \begin{minipage}{0.8in}
      \centerline{\includegraphics[width=1\textwidth]{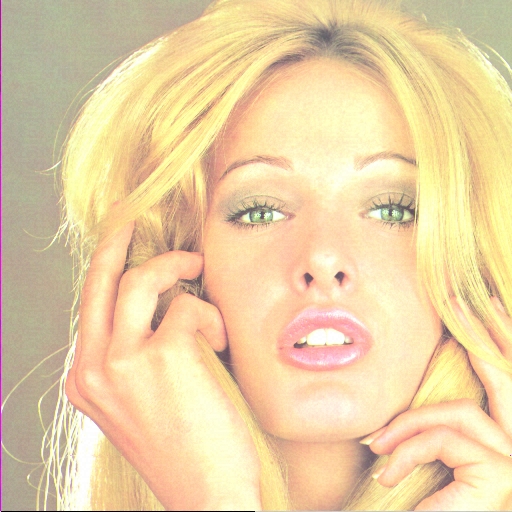}}
    \end{minipage}
    \hfill
    \begin{minipage}{0.8in}
      \centerline{\includegraphics[width=1\textwidth]{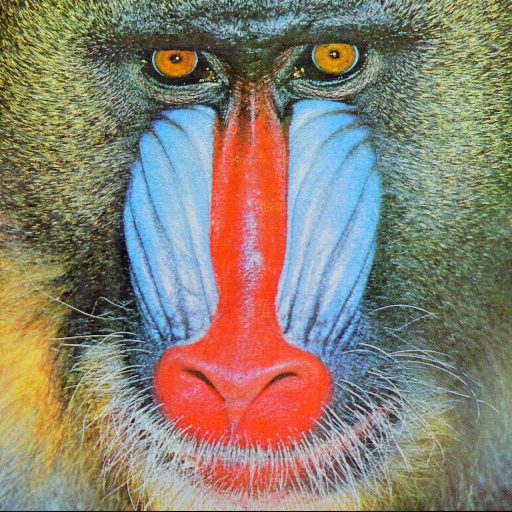}}
    \end{minipage}
    \hfill
    \begin{minipage}{0.8in}
      \centerline{\includegraphics[width=1\textwidth]{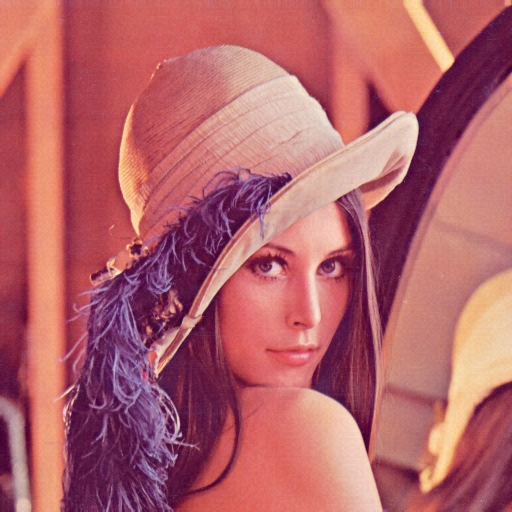}}
    \end{minipage}
    \hfill
    \begin{minipage}{0.8in}
      \centerline{\includegraphics[width=1\textwidth]{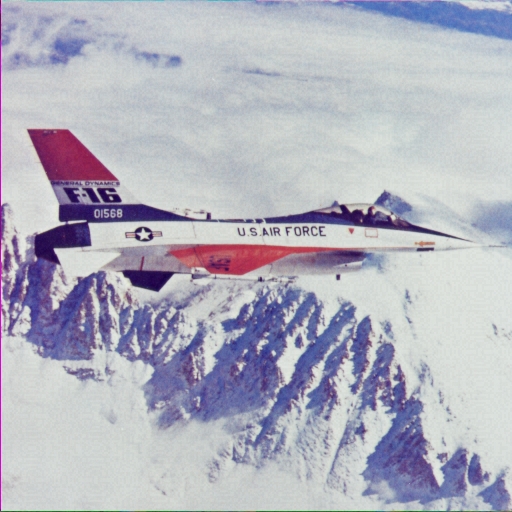}}
    \end{minipage}
    \hfill
    \begin{minipage}{0.8in}
      \centerline{\includegraphics[width=1\textwidth]{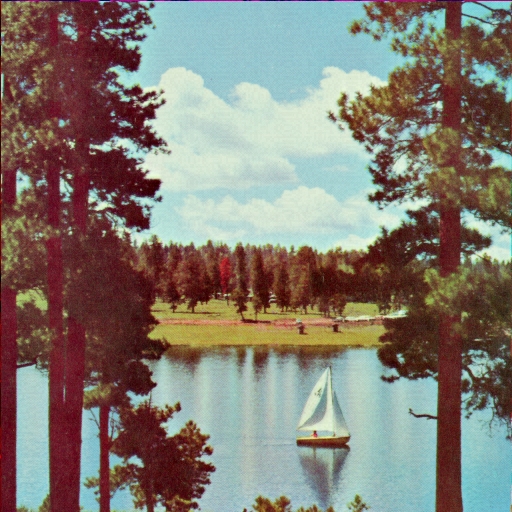}}
    \end{minipage}
    \hfill
    \begin{minipage}{0.8in}
      \centerline{\includegraphics[width=1\textwidth]{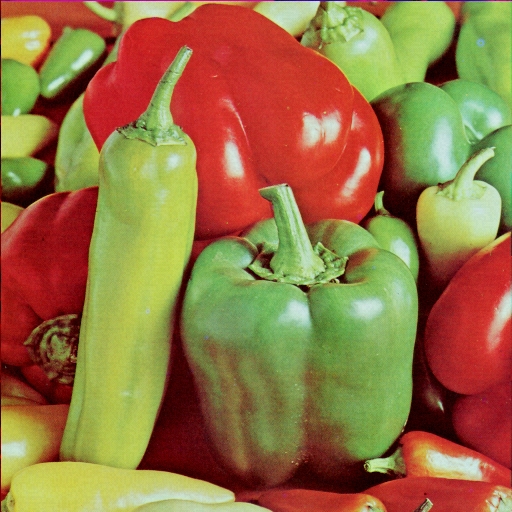}}
    \end{minipage}
    \hfill
    \begin{minipage}{0.8in}
      \centerline{\includegraphics[width=1\textwidth]{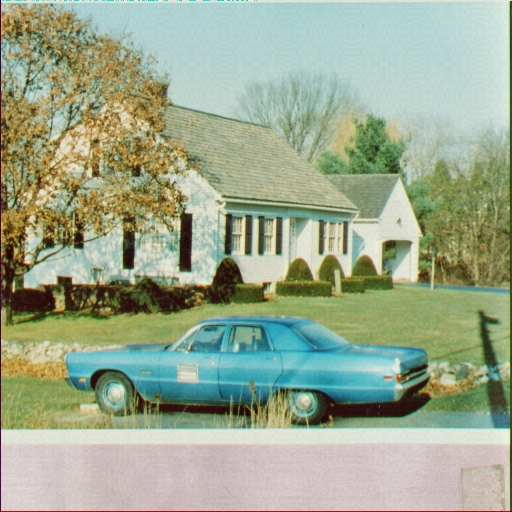}}
    \end{minipage}
    \caption{\emph{Top row:} Host images; \emph{Bottom row:} Watermarked images generated by the proposed scheme.}
    \label{host_image}
\end{figure*}

\section{Experimental Results}\label{experiment}
In this section, the proposed watermarking scheme based on symmetry is compared with some state-of-art watermarking schemes. The experimental parameters, setup of comparative experiment and test database are described in Section~\ref{setup}. In Section~\ref{RtoCIP}, \ref{RtoGD}, and \ref{RtoCA}, the robustness in different aspects is compared between the proposed scheme and other watermarking schemes.

\subsection{Implementation Details}\label{setup}
In our scheme, the size of spread-spectrum matrix $\boldsymbol{R}$ is an important parameter. Obviously, bigger $\boldsymbol{R}$ makes the watermark more robust to common image processing operations. Meanwhile, smaller $\boldsymbol{R}$ means more watermark units within the same size area, that is, there are more symmetrical peaks to better approximate the geometric distortions. In our experiment, we choose $L_{R}=4$ so the watermark is robust to various geometric distortions even RBA. And we choose $L_{m}=64$. To evaluate the performance between the proposed scheme and the state-of-art ones fairly, the message is embedded directly without ECC encoded and the bits error quantity (BEQ) is recorded.

As mentioned in Section~\ref{introduction}, only watermarking schemes belonging to the last two categories have the possibility that being robust to local geometric distortions. Voloshynovskiy \emph{et.al} \cite{Multibit} and Tian \emph{et.al} \cite{LDFT} respectively belong to these two categories and are resilient to local geometric distortions. In the rest categories, the robustness of Kang \emph{et.al} \cite{ULPM} to global geometric distortions and common image processing operations is one of the best. And we compare the proposed scheme with these three schemes. To illustrate the experimental results concisely, in the following part, we use V\_ACF \cite{Multibit}, T\_LDFT \cite{LDFT}, and K\_ULPM \cite{ULPM} to represent the three watermarking schemes above.

The test host images include eight colorful images, selected from USC-SIPI image database \cite{database1} and one hundred gray images, randomly selected from BOSSBase database \cite{database2}. Fig.~\ref{host_image} shows part of the host images and their corresponding watermarked images generated by the proposed scheme. The average peak signal-to-noise ratio (PSNR) of 108 watermarked images is 41.35dB. The robustness of the proposed scheme to distortions is evaluated by the average bits error quantity (BEQ) of watermarked images under the corresponding distortions. To other watermarking schemes in the comparative experiment, the same experimental setup and process are applied. As shown in Table~\ref{psnr}, the average PSNR values of watermarked images generated respectively by four different watermarking schemes are set to the same level of $41\pm0.4$dB. 

Most distortions in our experiment are realized by Stirmark 4.0 Benchmark \cite{stirmark1,stirmark2}.

\begin{table}[h]
\large
\caption{\\Average PNSR of Watermarked Images Generated by Different Schemes.} \centering
\label{psnr}
\scalebox{0.66}{
\begin{tabular}{c|cccc}
\toprule
\textbf{Scheme} & \textbf{V\_ACF} \cite{Multibit} & \textbf{T\_LDFT} \cite{LDFT} & \textbf{K\_ULPM} \cite{ULPM} & \textbf{Proposed} \\
\midrule
\textbf{PNSR(dB)} & 40.74 & 40.61 & 41.03 & 41.35 \\
\bottomrule
\end{tabular}
}
\end{table}

\subsection{Robustness to Common Image Processing Operations}\label{RtoCIP}
In this subsection, we compare the capability of V\_ACF \cite{Multibit}, T\_LDFT \cite{LDFT}, K\_ULPM \cite{ULPM}, and the proposed scheme to recover the hidden message under common image processing operations, including JPEG compression, Gauss noise, and average filtering. The experimental results are shown as follows.
\subsubsection{Robustness to JPEG Compression}
The watermarked images are compressed with JPEG compression with quality factor (QF) from 15 to 90. The average BEQ is listed in Table~\ref{JPEG}. The proposed scheme performs well under JPEG compression with all quality factors. To be specific, the performance of the proposed scheme is similar to K\_ULPM \cite{ULPM}'s and better than the rest schemes'.


\begin{table}[h]
\large
\caption{\\Average BEQ of Watermarked Images under JPEG Compression.} \centering
\label{JPEG}
\scalebox{0.63}{
\begin{tabular}{c|cccc}
\toprule
\textbf{Quality Factor} & \textbf{V\_ACF} \cite{Multibit} & \textbf{T\_LDFT} \cite{LDFT} & \textbf{K\_ULPM} \cite{ULPM} & \textbf{Proposed} \\
\midrule
90	&	0	&	6.111	&	0.037	&	\textbf{0}	\\
\midrule
80	&	0	&	16.352	&	0.343	&	\textbf{0}	\\
\midrule
70	&	0	&	20.740	&	0.482	&	\textbf{0}	\\
\midrule
60	&	0.046	&	22.815	&	0.630	&	\textbf{0}	\\
\midrule
50	&	0.213	&	24.102	&	1.278	&	\textbf{0.157}	\\
\midrule
40	&	0.657	&	26.963	&	1.954	&	\textbf{0.481}	\\
\midrule
30	&	\textbf{0.843}	&	29.926	&	1.676	&	1.741	\\
\midrule
20	&	4.704	&	32.176	&	\textbf{2.389}	&	4.657	\\
\midrule
15	&	7.037	&	31.278	&	\textbf{4.102}	&	6.046	\\
\bottomrule
\end{tabular}
}
\end{table}

\subsubsection{Robustness to Gauss Noise}
The watermarked images are corrupted by Gauss noise with the variance from 0.0001 to 0.01. As Table~\ref{Gauss} shows, the 64-bit message can be recovered with few errors by the proposed watermarking scheme. Meanwhile, V\_ACF \cite{Multibit} also performs well. The proposed watermarking scheme and V\_ACF are both spatial watermark and have a similar framework. Their low BEQ is a result of accumulation operation, which is a common spatial watermark enhancement method. The impact of Gauss noise could be excellently decreased by accumulating the repeated watermark units.

\subsubsection{Robustness to Average Filtering}
Average filters of different size are applied on the watermarked images and the average BEQ is recorded in Table~\ref{Average}. The proposed watermarking scheme could recover the message without error under $3\times3$ average filtering. When the filter size becoming bigger, the performance of the proposed scheme is worse than K\_ULPM \cite{ULPM}'s but better than the rest two schemes'.

\begin{table}[h]
\large
\caption{\\Average BEQ of Watermarked Images under Gauss Noise.} \centering
\label{Gauss}
\scalebox{0.63}{
\begin{tabular}{c|cccc}
\toprule
\textbf{Variance} & \textbf{V\_ACF} \cite{Multibit} & \textbf{T\_LDFT} \cite{LDFT} & \textbf{K\_ULPM} \cite{ULPM} & \textbf{Proposed} \\
\midrule
0.0001	&	0	&	2.824	&	1.380	&	\textbf{0}	\\
\midrule
0.001	&	0.185	&	27.491	&	1.435	&	\textbf{0.056}	\\
\midrule
0.01	&	0.796	&	31.759	&	2.380	&	\textbf{0.370}	\\
\bottomrule
\end{tabular}
}
\end{table}

\begin{table}[h]
\large
\caption{\\Average BEQ of Watermarked Images with Average Filtering.} \centering
\label{Average}
\scalebox{0.63}{
\begin{tabular}{c|cccc}
\toprule
\textbf{Filter Size} & \textbf{V\_ACF} \cite{Multibit} & \textbf{T\_LDFT} \cite{LDFT} & \textbf{K\_ULPM} \cite{ULPM} & \textbf{Proposed} \\
\midrule
$3 \times 3$	&	0.083	&	21.648	&	1.704	&	\textbf{0.028}	\\
\midrule
$5 \times 5$	&	14.750	&	28.278	&	\textbf{2.676}	&	13.657	\\
\midrule
$7 \times 7$	&	31.167	&	31.120	&	\textbf{8.500}	&	27.954	\\
\bottomrule
\end{tabular}
}
\end{table}

\subsection{Robustness to Geometric Distortions}\label{RtoGD}
The robustness of watermarking schemes to geometric distortions, including global and local geometric distortions, are investigated in this subsection. Besides rotation, scaling, translation, and cropping, affine transformation, lines removal, and aspect ratio change are also set as global geometric distortions in comparative experiments. Meanwhile, RBA proposed in Stirmark is set as local geometric distortion in the comparative experiment. There is no specific experiment to translation because translation is a basic geometric distortion and usually exists along with other geometric distortions, like rotation and cropping.

\subsubsection{Robustness to Rotation}
The watermarked images are rotated by some specific angles, including some small, precise angles, to test the robustness of watermarking schemes to rotation. The experimental results are listed in Table~\ref{rotation}. The proposed scheme almost recovers the message with very few errors from all of the rotation angles. And the rest schemes have higher average BEQ, especially under some specific rotation angles.

\begin{table}[h]
\large
\caption{\\Average BEQ of Rotated Watermarked Images.} \centering
\label{rotation}
\scalebox{0.63}{
\begin{tabular}{c|cccc}
\toprule
\textbf{Rotate Angle} & \textbf{V\_ACF} \cite{Multibit} & \textbf{T\_LDFT} \cite{LDFT} & \textbf{K\_ULPM} \cite{ULPM} & \textbf{Proposed} \\
\midrule
$0.25^\circ$	&	3.343	&	5.148	&	3.204	&	\textbf{0.093}	\\
\midrule
$0.5^\circ$	&	2.585	&	5.620	&	2.259	&	\textbf{0.019}	\\
\midrule
$1^\circ$	&	1.222	&	4.481	&	2.352	&	\textbf{0.046}	\\
\midrule
$5^\circ$	&	0.750	&	8.593	&	4.833	&	\textbf{0.046}	\\
\midrule
$30^\circ$	&	0.259	&	7.556	&	5.926	&	\textbf{0.139}	\\
\midrule
$45^\circ$	&	0.509	&	12.759	&	8.287	&	\textbf{0.157}	\\
\midrule
$90^\circ$	&	0.167	&	28.574	&	24.389	&	\textbf{0.065}	\\
\bottomrule
\end{tabular}
}
\end{table}

\subsubsection{Robustness to Scaling}
The width and height of watermarked images are scaled proportionally with the scaling factor from 0.5 to 2.0. As shown in Table~\ref{scaling}, under all scaling factors, the average BEQ of the proposed scheme is 0. Our scheme and V\_ACF \cite{Multibit} have similar performance and are much better than the rest.

\begin{table}[h]
\large
\caption{\\Average BEQ of Scaled Watermarked Images.} \centering
\label{scaling}
\scalebox{0.63}{
\begin{tabular}{c|cccc}
\toprule
\textbf{Scaling Factor} & \textbf{V\_ACF} \cite{Multibit} & \textbf{T\_LDFT} \cite{LDFT} & \textbf{K\_ULPM} \cite{ULPM} & \textbf{Proposed} \\
\midrule
0.5	&	0.083	&	24.537	&	11.787	&	\textbf{0}	\\
\midrule
0.75	&	0	&	11.574	&	4.935	&	\textbf{0}	\\
\midrule
0.9	&	0	&	6.473	&	3.139	&	\textbf{0}	\\
\midrule
1.1	&	0	&	8.019	&	3.056	&	\textbf{0}	\\
\midrule
1.5	&	0	&	20.343	&	6.315	&	\textbf{0}	\\
\midrule
2	&	0	&	26.167	&	18.796	&	\textbf{0}	\\
\bottomrule
\end{tabular}
}
\end{table}

\subsubsection{Robustness to Cropping}
The watermarked images are cropped by the ratio from $1\%$ to $75\%$. Both width and height are cropped. For example, if a watermarked image of size $512\times512$ is cropped $75\%$, only the central region of size $128\times128$ is reserved whose area is only $6.25\%$ of the original one. The experimental results are listed in Table~\ref{cropping}. The proposed watermarking scheme can recover the message almost without error when the cropping ratio is from $1\%$ to $50\%$. When the cropping ratio is $75\%$, the average BEQ of the proposed scheme is only 5.537, much less than the rest schemes'.


\begin{table}[h]
\large
\caption{\\Average BEQ of Cropped Watermarked Images.} \centering
\label{cropping}
\scalebox{0.63}{
\begin{tabular}{c|cccc}
\toprule
\textbf{Cropping Ratio} & \textbf{V\_ACF} \cite{Multibit} & \textbf{T\_LDFT} \cite{LDFT} & \textbf{K\_ULPM} \cite{ULPM} & \textbf{Proposed} \\
\midrule
$1\%$	&	0.093	&	0.231	&	3.917	&	\textbf{0}	\\
\midrule
$5\%$	&	0	&	1.157	&	4.833	&	\textbf{0}	\\
\midrule
$10\%$	&	0.269	&	1.083	&	5.546	&	\textbf{0}	\\
\midrule
$25\%$	&	0.657	&	3.000	&	8.111	&	\textbf{0}	\\
\midrule
$50\%$	&	0.694	&	5.444	&	17.963	&	\textbf{0}	\\
\midrule
$75\%$	&	9.824	&	12.778	&	28.009	&	\textbf{5.537}	\\
\bottomrule
\end{tabular}
}
\end{table}

\subsubsection{Robustness to Affine Transformation}
The watermarked images are transformed with different affine transformation matrix as follows:
\begin{equation}
    \left[
\begin{array}{c}
     x^{\prime}\\
     y^{\prime} 
\end{array}
    \right]=
    \left[
\begin{array}{cc}
     a&b\\
     c&d 
\end{array}
    \right]
    \left[
\begin{array}{c}
     x\\
     y 
\end{array}
    \right].
\end{equation}
By adjusting $a$, $b$, $c$, and $d$, X-shearing, Y-shearing, and XY-shearing of different strengths would be applied to watermarked images. According to the experimental results in Table~\ref{affine}, under all the tested affine transformations, the BEQ of the message recovered by the proposed scheme is 0. Meanwhile, under a specific affine transformation, the BEQ of V\_ACF \cite{Multibit} is 0.407, the BEQ of T\_LDFT \cite{LDFT} is 8.269, and the BEQ of K\_ULPM \cite{ULPM} is up to 30.870.


\begin{table}[h]
\large
\caption{\\Average BEQ of Affine Transformed Watermarked Images.} \centering
\label{affine}
\scalebox{0.55}{
\begin{tabular}{c|cccc}
\toprule
\begin{array}{c}
     \textbf{Affine Transform Matrix} \\
     \emph{a \quad b \quad c \quad d}
\end{array} & \textbf{V\_ACF} \cite{Multibit} & \textbf{T\_LDFT} \cite{LDFT} & \textbf{K\_ULPM} \cite{ULPM} & \textbf{Proposed} \\
\midrule
1 \quad 0 \quad 0.01 \quad 1&	0	&	1.139	&	5.556	&	\textbf{0}	\\
\midrule
1 \quad 0 \quad 0.05 \quad 1&	0	&	3.870	&	28.148	&	\textbf{0}	\\
\midrule
1 \quad 0.01 \quad 0 \quad 1&	0	&	2.583	&	6.861 &	\textbf{0}	\\
\midrule
1 \quad 0.05 \quad 0 \quad 1&	0.046	&	2.269	&	30.917	&	\textbf{0}	\\
\midrule
1 \; 0.01 \; 0.01 \; 1&	0.185	&	5.806	&	11.435	&	\textbf{0}	\\
\midrule
1 \; 0.05 \; 0.05 \; 1&	0.407	&	8.269	&	30.870	&	\textbf{0}	\\
\bottomrule
\end{tabular}
}
\end{table}

\subsubsection{Robustness to Aspect Ratio Change}
The width and height of watermarked images are scaled respectively with different scaling factors. In Table~\ref{aspect}, the height of watermarked images is scaled with the first scaling factor and the width is scaled with the second one. As Table~\ref{aspect} shows, the proposed watermarking scheme can recover the embedded message without error with all aspect ratio change. V\_ACF \cite{Multibit} performs as well as the proposed scheme. The average BEQ of T\_LDFT \cite{LDFT} is over 9, and the average BEQ of K\_ULPM \cite{ULPM} is over 31.

\begin{table}[h]
\large
\caption{\\Average BEQ of Disproportionally Scaled Watermarked Images.} \centering
\label{aspect}
\scalebox{0.63}{
\begin{tabular}{c|cccc}
\toprule
\textbf{Scaling Factors} & \textbf{V\_ACF} \cite{Multibit} & \textbf{T\_LDFT} \cite{LDFT} & \textbf{K\_ULPM} \cite{ULPM} & \textbf{Proposed} \\
\midrule
$0.9\times1.1$	&	0	&	9.630	&	32.407	&	\textbf{0}	\\
\midrule
$1.5\times0.8$	&	0	&	21.352	&	31.083	&	\textbf{0}	\\
\midrule
$0.7\times1.8$	&	0	&	25.519	&	32.271	&	\textbf{0}	\\
\bottomrule
\end{tabular}
}
\end{table}

\subsubsection{Robustness to Lines Removal}
To investigate the robustness to lines removal, $1\%$, $5\%$ and $10\%$ lines in row and column of watermarked images are removed. These removed lines are distributed at regular intervals. As Table~\ref{RML} shows, under all tested lines removal, the proposed scheme has BEQ of 0, much lower than the BEQ of the rest schemes.

\begin{table}[h]
\large
\caption{\\Average BEQ of Watermarked Images under Lines Removal.} \centering
\label{RML}
\scalebox{0.58}{
\begin{tabular}{c|cccc}
\toprule
\textbf{Lines Removal Ratio} & \textbf{V\_ACF} \cite{Multibit} & \textbf{T\_LDFT} \cite{LDFT} & \textbf{K\_ULPM} \cite{ULPM} & \textbf{Proposed} \\
\midrule
$1\%$	&	0	&	0.426	&	5.907	&	\textbf{0}	\\
\midrule
$5\%$	&	0.056	&	1.241	&	4.954	&	\textbf{0}	\\
\midrule
$10\%$	&	0.083	&	4.185	&	6.500	&	\textbf{0}	\\
\bottomrule
\end{tabular}
}
\end{table}

\subsubsection{Robustness to Random Bending Attack}
RBA of Stirmark is one of the most typical local geometric distortions. Considering the randomness of RBA, to specific attack strength, each watermarked image is attacked 5 times to get the average performance. Thus, each average BEQ in Table~\ref{table_RBA} is calculated from the messages recovered from $108\times 5$ watermarked images. The experimental results with varying strengths are shown in Table~\ref{table_RBA} and the average BEQ of the proposed watermarking scheme is no more than 0.7. The proposed watermarking scheme has excellent performance under RBA, even better than the performance of T\_LDFT \cite{LDFT} which is based on local geometrically invariant features.

Besides, the proposed watermarking scheme has a better performance compared with V\_ACF \cite{Multibit}, especially under the RBA with big strength. As we have mentioned in Section.~\ref{illustration_of_SP}, the symmetrical peaks in the proposed scheme are resilient to perspective transformation and perspective transformation can better approximate local geometric distortions. The parameters of the perspective transformations can be obtained directly from the watermark units map $\boldsymbol{M}$, just like Fig.~\ref{RST_map}(d).

\begin{table}[h]
\large
\caption{\\Average BEQ of Watermarked Images under RBA.} \centering
\label{table_RBA}
\scalebox{0.6}{
\begin{tabular}{c|cccc}
\toprule
\textbf{Strength of RBA} & \textbf{V\_ACF} \cite{Multibit} & \textbf{T\_LDFT} \cite{LDFT} & \textbf{K\_ULPM} \cite{ULPM} & \textbf{Proposed} \\
\midrule
0.1	&	0	&	5.113	&	1.370	&	\textbf{0}	\\
\midrule
0.2	&	0	&	5.391	&	2.243	&	\textbf{0}	\\
\midrule
0.3	&	0.074	&	5.406	&	6.133	&	\textbf{0}	\\
\midrule
0.4	&	0.065	&	5.957	&	9.943	&	\textbf{0.056}	\\
\midrule
0.5	&	0.248	&	5.576	&	12.061	&	\textbf{0.124}	\\
\midrule
0.6	&	2.764	&	5.254	&	13.235	&	\textbf{0.278}	\\
\midrule
0.7	&	6.191	&	5.822	&	15.017	&	\textbf{0.472}	\\
\midrule
0.8	&	5.687	&	5.843	&	17.170	&	\textbf{0.659}	\\
\midrule
0.9	&	6.185	&	5.791	&	20.044	&	\textbf{0.674}	\\
\midrule
1.0	&	9.369	&	5.828	&	22.650	&	\textbf{0.602}	\\
\bottomrule
\end{tabular}
}
\end{table}


\begin{table}[h]
\large
\caption{\\Average BEQ of Watermarked Images under Combined Attacks.} \centering
\label{CA}
\scalebox{0.55}{
\begin{tabular}{l|cccc}
\toprule
\textbf{Combined Attack Types} & \textbf{V\_ACF} \cite{Multibit} & \textbf{T\_LDFT} \cite{LDFT} & \textbf{K\_ULPM} \cite{ULPM} & \textbf{Proposed} \\
\midrule
Average\, +\, Gauss 	&	0.111	&	29.519	&	1.824	&	\textbf{0.083}	\\
									
JPEG\, + \, Average 	&	0.083 &	26.130	&	1.639	&	\textbf{0.021}	\\
									
JPEG\, + \, Gauss 	&	0	&	28.287	&	2.102	&	\textbf{0}	\\
\midrule									
Cropping\, + \, Rotation 	&	1.120	&	5.556	&	11.963	&	\textbf{0.157}	\\
									
Cropping\, + \, Scaling 	&	0.713	&	22.167	&	9.972	&	\textbf{0}	\\
									
Scaling\, + \, Rotation 	&	0.611	&	18.972	&	7.963	&	\textbf{0.222}	\\
\midrule									
Cropping\, + \, Average 	&	1.157	&	25.806	&	8.491	&	\textbf{0.044}	\\
									
Cropping\, + \, Gauss 	&	0.935	&	24.426	&	8.148	&	\textbf{0}	\\
									
JPEG\, + \, Cropping 	&	1.111	&	23.917	&	8.796	&	\textbf{0}	\\
									
JPEG\, + \, Rotation 	&	0.731	&	24.657	&	5.963	&	\textbf{0.231}	\\
									
JPEG\, + \, Scaling 	&	0	&	26.037	&	5.852	&	\textbf{0}	\\
									
Rotation\, + \, Average 	&	0.451	&	24.935	&	6.685	&	\textbf{0.204}	\\
									
Rotation\, + \, Gauss 	&	0.583	&	30.759	&	5.750	&	\textbf{0.185}	\\
									
Scaling\, + \, Average 	&	0.241	&	28.398	&	5.657	&	\textbf{0.093}	\\
									
Scaling\, + \, Gauss 	&	0	&	32.176	&	5.713	&	\textbf{0}	\\
\midrule									
RBA\, + \, Average 	&	0.102	&	25.037	&	10.278	&	\textbf{0.102}	\\
									
RBA\, + \, Cropping 	&	0.685	&	8.046	&	28.102	&	\textbf{0.074}	\\
									
RBA\, + \, Gauss 	&	0.111	&	31.704	&	10.324	&	\textbf{0.009}	\\
									
RBA\, + \, JPEG 	&	0.185	&	24.954	&	10.037	&	\textbf{0.102}	\\
									
RBA\, + \, Rotation 	&	0.343	&	16.444	&	3.389	&	\textbf{0.157}	\\
									
RBA\, + \, Scaling 	&	0.194	&	22.796	&	6.769	&	\textbf{0.019}	\\
\bottomrule
\end{tabular}
}
\end{table}

\subsection{Robustness to Combined Attacks}\label{RtoCA}
In this subsection, we select a couple of representative single attacks, that is, JPEG compression with QF 70, Gauss noise with variance 0.001, $3\times3$ average filtering, rotating $30^{\circ}$, scaling with factor 0.75, cropping $25\%$, and RBA with strength 0.3. One combined attack is generated by combining two of these single attacks. The experimental results are listed in Table~\ref{CA}, where parameters of attacks are omitted. The average BEQ of the proposed scheme is less than 1, much lower than the rest schemes. It is not a surprising result because the proposed scheme, different from others, has better performance under all these single attacks.

\section{Conclusion}\label{conclusion}
This paper presents a blind and robust watermarking scheme, including a novel watermark synchronization process. The proposed synchronization process is based on the symmetry of the symmetrical watermark, which has several advantages comparing to prior similar schemes. Besides, we propose to minimize the mean-square error to get the watermark estimation, use auto-convolution function to quickly calculate the symmetry, and apply hypothesis testing to determine the watermark state. We believe that the proposed watermark synchronization process based on symmetry has the potential to become an improved version of the one based on periodicity and help similar schemes to improve their performance. According to the experimental results, the proposed watermarking scheme has excellent performance under various distortions, including RBA, global geometric distortions, common image processing operations, and combined attacks. In further work, we will focus on the application of the proposed watermarking scheme on the print-camera process.

\ifCLASSOPTIONcaptionsoff
  \newpage
\fi

\bibliographystyle{IEEEtran}
\bibliography{IEEEabrv,Bibliography}

\vfill

\end{document}